\documentclass[onecolumn]{aastex7}
\usepackage{hyperref}
\hypersetup{pdfnewwindow=true,colorlinks=true,linkcolor=xlinkcolor,citecolor=xlinkcolor,
filecolor=xlinkcolor,urlcolor=xlinkcolor,final=true}
\usepackage{bookmark}

\newcounter{minirefcount}
\newcommand{\mr}[2]{\refstepcounter{minirefcount}\label{#2}(\arabic{minirefcount}) #1}

\begin{document}

\title{A Comprehensive Analysis of Rovibrational CO in the Era of JWST}

\author[0000-0002-4555-5144]{D. Annie Dickson-Vandervelde}
\email{dadicksonvandervelde@gmail.com}
\affiliation{Department of Physics and Astronomy, Vassar College, 124 Raymond Avenue, Poughkeepsie, NY 12604, USA}

\author[0000-0003-3682-6632]{Colette Salyk}
\email{}
\affiliation{Department of Physics and Astronomy, Vassar College, 124 Raymond Avenue, Poughkeepsie, NY 12604, USA}

\author[0000-0003-0787-1610]{Geoffrey A. Blake}
\email{}
\affiliation{Division of Geological and Planetary Sciences, California Institute of Technology, MC 150-21, Pasadena, CA 91125, USA}

\author[0009-0009-9795-2879]{Clara Ross}
\email{}
\affiliation{Center for Astrophysics, Harvard \& Smithsonian, 60 Garden Street, Cambridge, MA 02138-1516, USA}

\author[0000-0001-9344-0096]{Adwin Boogert}
\email{}
\affiliation{Institute for Astronomy, University of Hawai’i at Manoa, 2680 Woodlawn Drive, Honolulu, HI 96822, USA}

\author[0000-0001-7552-1562]{Klaus Pontoppidan}
\email{}
\affiliation{Jet Propulsion Laboratory, California Institute of Technology, 4800 Oak Grove Drive, Pasadena, CA 91109, USA}

\begin{abstract}
    We present an analysis of CO rovibrational emission lines in the 183 infrared spectra of nearby Class II objects obtained with the NIRSPEC instrument on the Keck II telescope over the past two decades. The sample includes a broad range of stellar mass (both T Tauri and Herbig Ae/Be) and disk evolutionary states (from full to debris disks). We find that 53\% of the sample has CO rovibrational emission lines present in their spectrum with disk/stellar subtype detection rates of 82\% for transition disks, 61\% for Herbigs, and 77\% for CTTSs. Although there is no discernible difference between T Tauri and Herbig Ae/Be star CO detection rates, the detection of accretion and of CO are statistically correlated in T Tauri stars but not in Herbig Ae/Be objects. Within the sample of T Tauri stars, we find that no weak-line T Tauri stars have CO rovibrational emission lines. We use slab modeling to analyze the density, temperature, and emitting area of the sample. The retrieval results imply that Herbig Ae/Be objects tend to have cooler and larger CO emitting regions than T Tauri stars. We find that the CO emitting area is not a thin ring as defined by temperature, but a ring of varying size likely dependent on the structure of the disk. 
    We also present guidelines on how to approach CO rovibrational emission lines in JWST spectra and present methods for linking ground-based observations with JWST spectra. This includes line-to-continuum ratio estimates based on stellar mass and accretion rate.
\end{abstract}

\section{Introduction}\label{sec:Intro}

Tracing the gas located in the inner few AU of disks is critical for understanding the conditions of terrestrial planet formation. Terrestrial planets most likely form in the innermost regions of protoplanetary disks, where the material begins to become an interface to the host star. 
Spatially resolving these regions is difficult, even with current adaptive optics (AO) systems (e.g., 1 au = 0.007$\arcsec$ at 150 pc), unlike the outer regions of disks which have been well studied via ALMA using dust continuum and cold gas tracers \citep{Andrews2020}, and in scattered light imaging in the near infrared (NIR) \citep{Benisty2023}. Since the early 2000s \citep[e.g.][]{Najita2003}, NIR spectroscopy of disks -- and specifically emission lines from the abundant molecule CO -- has been a key tool in understanding the bulk gas of this region. At temperatures from a few hundred to a few thousand K CO molecules emit in the infrared via rovibrational excitation. 

This rovibrational emission from CO primarily arises in the disk atmosphere, where temperature inversion (potentially assisted by gas-dust thermal decoupling and dust settling, \citealp{Kamp2004}) in so-called ``passive disks'' increases gas emission line brightness above the otherwise optically thick continuum \citep{ChiangGoldreich1997}. Such emission can also arise from the optically thin regions of cleared disks \citep{Carr2001,Salyk2009}. The excitation parameters required for these conditions, retrieved from the CO rovibrational emission, along with the line shape information provided by high resolution spectroscopy, allow the observer of these lines to obtain information about the inner few au of disks \citep{Banzatti2022}. This region is likely farther out than the accretion tracers (like H I) and is therefore complementary for studying this inner region.  Because the most abundant gas in the inner disk, H$_2$, remains challenging to detect \citep{Bitner2008}, CO serves as a proxy for the bulk gas within disks. It also serves as a ``denominator'' in disk chemical studies, i.e., researchers may report X/CO ratios instead of abundances relative to H$_2$ \citep{Pontoppidan2014}, as a more directly observable quantity.

Since the discovery of CO rovibrational emission in protoplanetary disks \citep{Carr2001} several works have explored the properties of inner disk gas, using a variety of IR spectrographs \citep[e.g., Keck-NIRSPEC, VLT-CRIRES, and IRTF-ISHELL,][]{NIRSPEC,CRIRES,iSHELL}. This emission probes two reservoirs of gas: emission from gas at the edge of the inner disk -- likely near the dust sublimation radius for most disks \citep{Salyk2011b} or at the edge of an inner clearing around Herbig Ae/Be stars (hereafter Herbigs) \citep{Banzatti2018} -- or the product of resonant scattering emission from radii as far out as 100 AU \citep{BlakeBoogert2004}. 
Detailed analysis of lineshapes with the R~$\sim$100,000 CRIRES instrument showed that emission lines can have multiple velocity contributions, including contributions from a disk wind \citep{Bast2011} or, in younger sources, outflows \citep{Herczeg2011}. Low energy transitions can also include absorption from foreground gas. In transition disks, the rovibrational emission probes the gas content in clearings \citep{Salyk2009,Pontoppidan2011} and helps constrain possible disk-planet interactions \citep{Brittain2009,Jensen2021}.  

In this work, we compile and present over two decades of spectra obtained with Keck-NIRSPEC and analyze the dataset as a whole. Past studies have investigated particular samples of objects (e.g., Class I, Herbigs, transitional) but have not looked collectively at CO gas in disks as a full sample.  In addition, studies with CRIRES had the advantage of a particularly high spectral resolution (R $\sim$ 45, $-$ 90,000), but the concomitant disadvantage of less spectral coverage, which elucidated more about the gas kinematics rather than the bulk gas properties \citep{Brown2013}. This work presents a comprehensive sample of 183 Class II and III disks and trends seen in their observed CO emission lines. 

Revisiting these datasets is particularly important in light of the recent launch of the James Webb Space Telescope (JWST) and its accompanying Integral Field Unit spectrographs, NIRSpec (hereafter JWST-NIRSpec, to distinguish it from Keck-NIRSPEC; \citealp{JWSTNIRSPEC}) and MIRI-MRS \citep{MIRIMRS}.  These spectrographs both include CO rovibrational emission (JWST-NIRSpec up to 5.29 $\mu$m or P(54), MIRI down to 4.91 $\mu$m, or P(25)), but will marginally spectrally resolve rovibrational emission for most disks \citep{Banzatti2024}.  A comprehensive understanding of existing high-resolution datasets will aid the analysis of the JWST datasets via resolved lineshapes (gas location) and broad ranges of excitation levels (slab modeling).  In addition, MIRI-MRS is being used to study disk chemistry; an understanding of the bulk gas content and disk atmospheric structure will aid in converting observed molecular spectra to chemical abundance profiles. Combining these datasets with JWST (e.g. \citealp{Temmink2024a}) will also present opportunities for studying line profile and intensity variability.

This paper is organized as follows: in Section~\ref{sec:ObsMeth}, we review the data collected, the reduction process, and the flux calculations for the entire sample, along with the biases present within this sample. We review rovibrational CO emission and our detection statistics in Section~\ref{sec:Analysis}. We present line fluxes and slab modeling results in Section~\ref{sec:Results} and discuss what this sample reveals about how CO compares to other disk tracers in Section~\ref{sec:Discussion}. Finally, we discuss the key takeaways for JWST in Section~\ref{sec:JWST}.

\begin{figure}
    \centering
    \includegraphics[width=\linewidth]{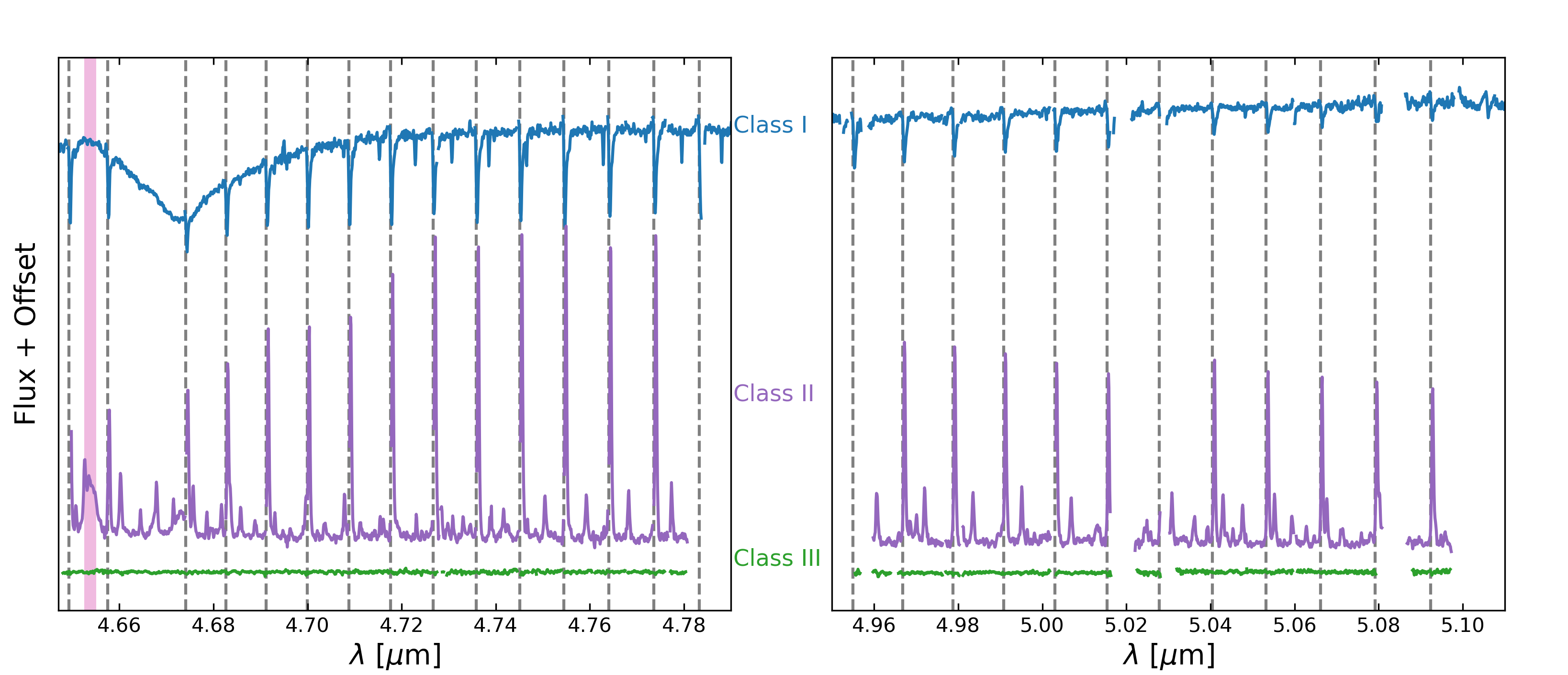}
    \caption{Three example Keck-NIRSPEC spectra of the different stellar classes of Pre-Main Sequence (PMS) objects: Class I (blue, LDN 1489), II (purple, DR Tau), and III (green, DI Tau). These spectra include two of the echelle orders from the ``M-wide'' filter. The dashed gray lines represent rovibrational CO(1--0) line wavelengths from HITRAN \citep{Gordon2022} and the shaded pink region represents the location of the H I Pfund $\beta$ transition. The additional emission/absorption lines visible adjacent to the marked CO(1--0) transitions are from CO(2--1) and $^{13}$CO(1--0) transitions.}
    \label{fig:ClassSpec}
\end{figure}

\section{Observations}\label{sec:ObsMeth}

Our spectra were obtained with Keck-NIRSPEC \citep{McLean1998}, a high-resolution infrared spectrometer on the Keck II telescope. The observations derive from observing runs spanning from 2001-2019 and form a large Keck-NIRSPEC survey of protoplanetary disks. Portions of this dataset have previously been presented, \citep{BlakeBoogert2004,Salyk2007,Salyk2009,Salyk2011b,Salyk2013}, and additional details about the observations and data reduction can be found therein.  Nevertheless, we provide a brief description here.

Spectra were obtained in the ``M-wide'' (4.420 $-$ 5.530 $\mu$m) filter in the cross-dispersed echelle mode.  Prior to 2016, Keck-NIRSPEC was not AO-fed, and most observations in our survey were performed with the 0$\arcsec$.43 $\times$ 24$\arcsec$ slit \citep[e.g.][]{Salyk2011b}, which represents a compromise between throughput and spectral resolution.  This mode provides a spectral resolving power of $R\sim25,000$ ($\Delta v\sim12.5$ km s$^{-1}$). Subsequent to an optics upgrade in 2016 \citep{Adkins2016}, Keck-NIRSPEC could be AO-fed; observations were then primarily performed using the 0.027 $\times$ 2.26 slit. Further improvements in 2018 resulted in an enhancement of instrumental resolution to $R\sim35,000$ ($\Delta v\sim9$ km s$^{-1}$), driven by the smaller pixel pitch of the new detector.

M-band observations include substantial sky and telescope thermal background emission, so targets were observed in ABBA nod patterns for background subtraction purposes.  Total image exposure times were typically limited to 60 seconds to minimize sky variability between adjacent exposures.  Integration times and coadds were then adjusted for each source to maintain detector linearity.  

Observations were typically performed in a set of two grating settings, although for a few targets additional settings were included. For the full data set, each setting has two usable orders; the resulting wavelength coverage can be seen in Figure \ref{fig:ClassSpec} (data after 2018 also include an order near 5.3-5.4 $\mu$m).  Extending from 4.65--5.15 $\mu$m with a gap in the middle, the settings cover R(0), R(1), P(1) $-$ P(12) and P(30) $-$ P(40) of the CO fundamental (v $=$ 1$-$0) band, along with the accretion tracers H I Pf $\beta$ and Hu $\varepsilon$ \citep{Salyk2013}.

All spectra from this work are available for viewing at the {\tt SpExoDisks} Database at spexodisks.com. 

\subsection{Data Reduction and Flux Calculation}

\begin{figure}
    \centering
    \includegraphics[width=0.85\linewidth]{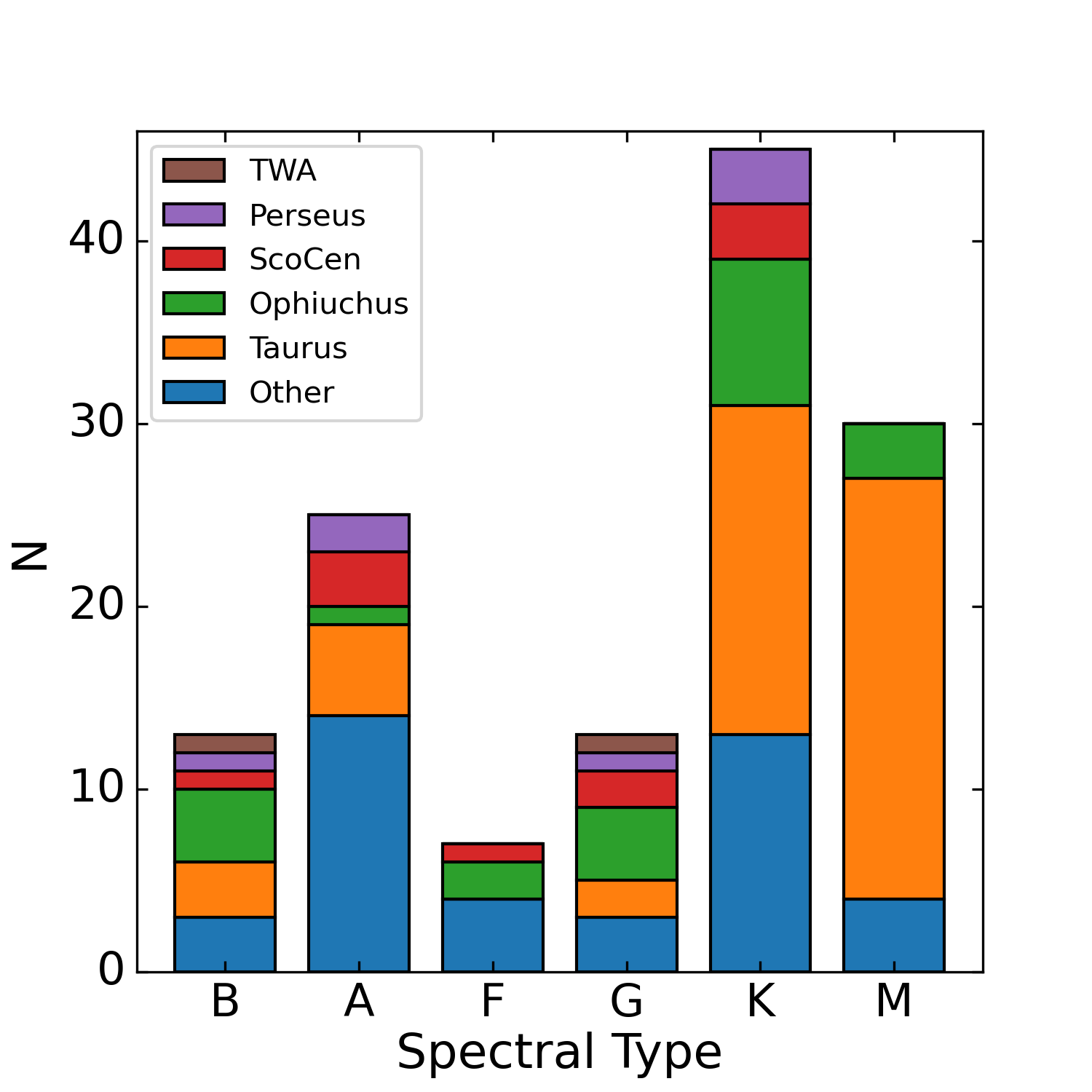}
    \caption{Histogram of the total number of objects in each spectral class color-coded by which star forming region or group they are associated with. The majority of our targets come from Taurus or Ophiuchus. The ``Other'' classification includes regions or groups with only a handful of stars that meet our detection criteria, or objects with no known, or conflicting, region membership in the literature.}
    \label{fig:Coverage}
\end{figure}

Data reduction was performed with our own tools, originally written in {\tt IDL}, but recently updated to {\tt Python}\footnote{available at https://github.com/Keck-DataReductionPipelines/NIRSPEC-Data-Reduction-Pipeline}, with an additional upgrade post the instrument upgrade in 2018. In these routines, A/B nod pairs are subtracted to remove background emission, and then the A-B images are rectified/linearized using source and telluric emission. Spectral extraction is performed with a width of customizable size, and resulting 1D spectra are averaged together. During this process any residual sky emission is subtracted using regions along the slit without stellar emission.  Wavelength calibration is performed using polynomial fits to telluric emission lines. Finally, spectra are divided by telluric standards --- observations of relatively featureless A and B stars, adjusted for small airmass differences --- after continuum and line absorption were removed from the standards based on model atmospheres and stellar rotation was accounted for using a boxcar smoothing function. The division process removes telluric absorption, but some errors remain in regions of high telluric absorption.  These regions are masked in the final analysis.  Observations obtained at different times of year (and therefore different Earth-induced Doppler shifts) are then combined together to ``fill in'' the line profiles.  In this work, final combined spectra are calibrated to a heliocentric reference frame. 

The spectra are flux calibrated using literature telluric standards. Line fluxes utilized in our analyses were obtained using the python package spectools\_ir \citep{spectools} --- specifically the flux\_calculator module.  This module uses the HITRAN database \citep{HITRAN} API via astroquery \citep{Ginsburg2019} to extract CO transition information, then performs Gaussian fits to generate both Gaussian-based and numerically integrated line fluxes. The numerically integrated line fluxes are preferable when the lineshape is more complicated, e.g. triangular.

A more detailed overview of NIRSPEC spectra acquisition methods used in this paper can be seen in \citet{Salyk2009}.

\subsection{The Sample}\label{subsec:Sample}

The sample for this paper is made up of Keck-NIRSPEC spectra of 183 objects. The majority of the sample (151 objects) have confirmed SED disk classifications; 120 Class II, 12 Class III, 12 debris disk, and 17 transition disk sources. While there has been debate over what a transition disk means in the age of mm-wave imaging (see \citealp{vanderMarel2023} for an overview), we define a transition disk as having large dust cavities and gaps that are identified via their infrared SED, not those simply with a mm-dust inner dust gap as imaged with ALMA. The sample covers a wide range of stellar and disk properties; see Table~\ref{table:Sample} for literature properties of our sample. 

The sample features two stellar types, T Tauri stars (TTSs) and Herbigs, for which each has two subtypes. Classified by accretion rates traced via the equivalent width of H$\alpha$ emission, weak-line T Tauri stars (WTTSs) are classified by lower accretion rates than  classical T Tauri stars (CTTSs) \citep{Feigelson1983,Wichmann1996}. Herbigs are divided into two Meeus groupings: Group I -- Herbigs with strong FIR emission compared to their MIR emission and considered to be flared -- and Group II -- sources with weak FIR emission compared to their MIR emission and considered to be flat, or dust-settled, disks \citep{Meeus2001,Bosman2019}. 
While the stellar subtypes for each star in the sample are sourced from a variety of large projects, each paper is noted in the reference column of Table~\ref{table:Sample} and their methods can be referenced therein.

\begin{figure}
    \centering
    \includegraphics[width=.38\linewidth]{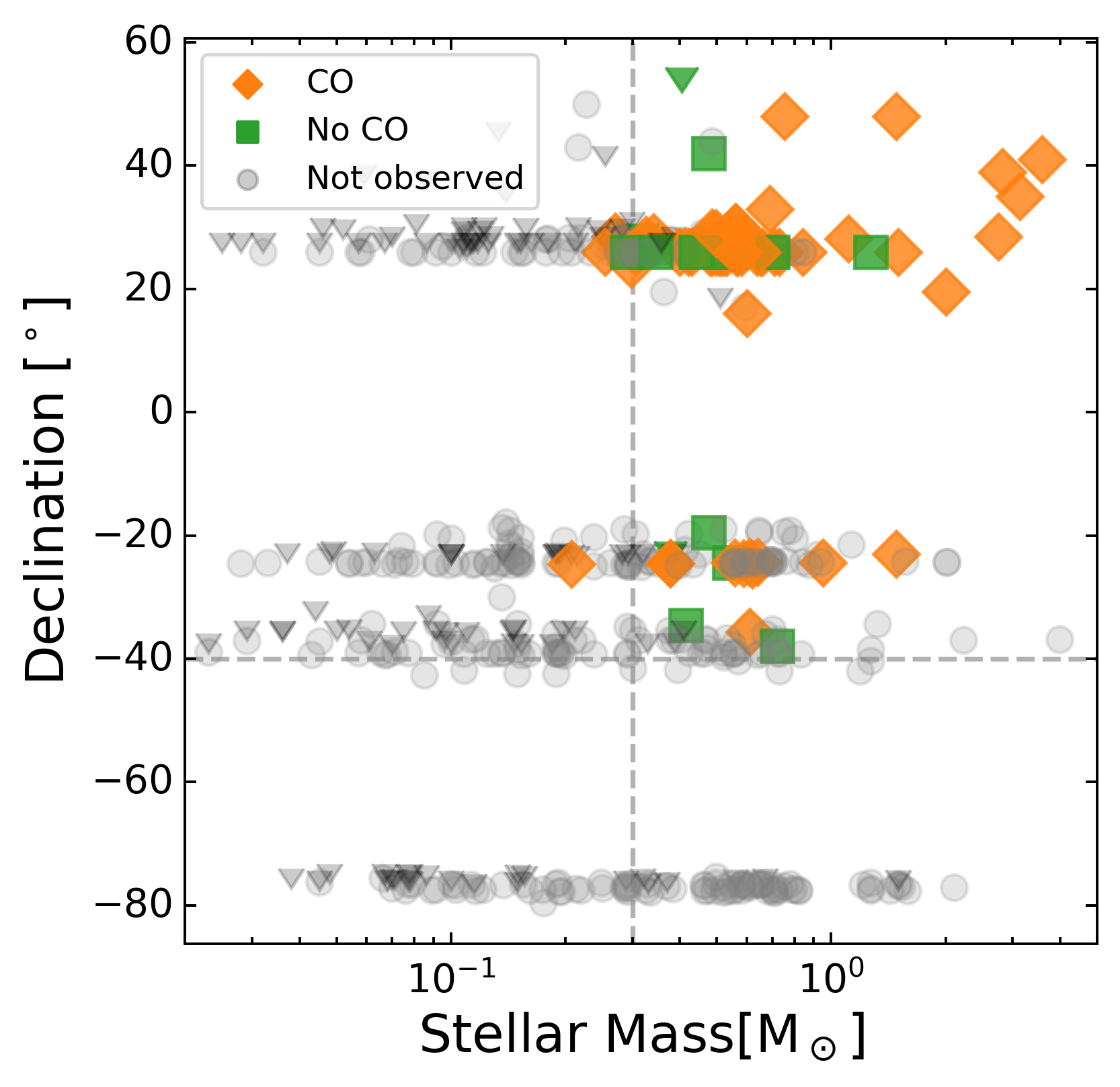}
    \includegraphics[width=.6\linewidth]{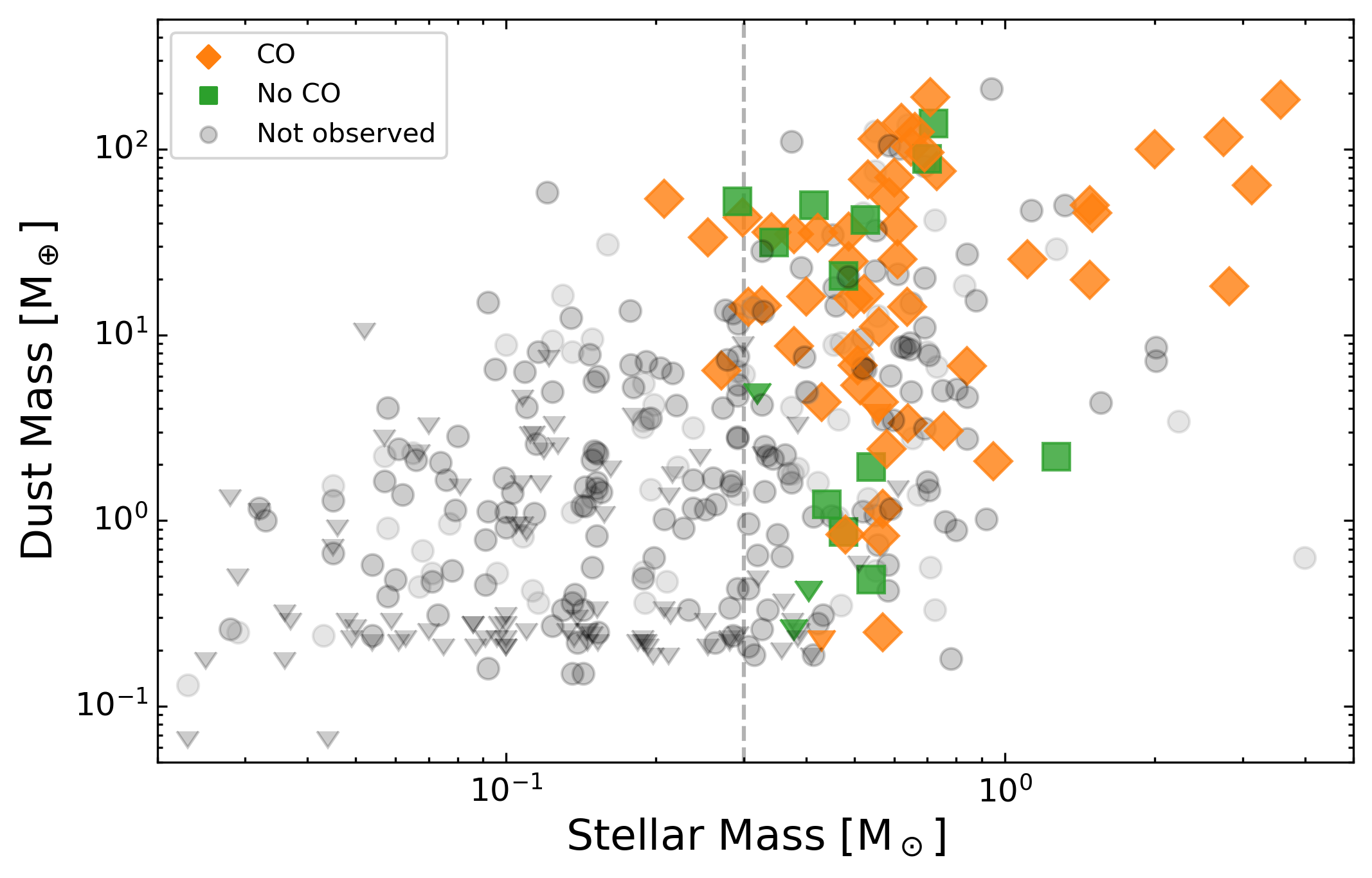}
    \caption{Left: The collected sample from \citet{Manara2023} with declination plotted versus stellar mass; upside triangles represent objects with only an upper limit on the associated dust mass. Objects that are also within our sample are plotted with orange diamonds if we detected CO and green squares if no CO was detected. The vertical dashed line represents a stellar mass of 0.3 M$_{\odot}$ and the horizontal dashed line represents a declination of $-40$. Right: The same sample as the left figure, but only for sources with declination $> -40$, that presents dust mass as a function of stellar mass. The lighter gray circles represent objects at the edge of Keck's observation limit ($-35 ^{\circ}$ to -40$^{\circ}$ Dec). }
    \label{fig:Manara}
\end{figure}

The bulk of this sample was chosen principally for its link to \textit{Spitzer} spectroscopy, especially the c2d program \citep{c2d2003} and contains a large number of sources from various regions within the Keck declination coverage ($_>\atop{^\sim}$ -40$^{\circ}$). The spectral type coverage of the sample, as shown in Figure~\ref{fig:Coverage}, has two main populations: A-type stars (i.e. the bulk of Herbigs) and K- and M-type stars (i.e. the bulk of T Tauri stars). The high coverage of Herbigs -- which mostly belong to Dark Clouds and OB Associations -- due to their brightness making these targets easier to observe, leads to an over-representation of the A spectral type in our sample. For the different associations in our sample, the selection tends to cover broadly the whole range of spectral types within a group. 

Using the \citet{Manara2023} (henceforth M23) sample, which combines disk surveys that are informed by the \textit{Spitzer} disk identifications to represent an estimated $>$ 80\% completeness of the total local disk population, we compare the portion of our sample that overlaps sufficiently with theirs in Figure~\ref{fig:Manara}. Considering only objects that would be observable from Keck (Dec $_>\atop{^\sim}$-40$^{\circ}$), our sample represents 11.6\% of M23. The sample representation increases to 47.0\% if we only consider objects with masses $>$ 0.3 M$_{\odot}$; smaller mass stars are not observable with Keck-NIRSPEC due to brightness limitations. The sample representation again increases, to 58.1\%, if we add a declination limit of $_>\atop{^\sim}$-35$^{\circ}$, since objects below this declination present additional challenges to observing with Keck and thus only a few were observed in our sample.

We exclude Keck-NIRSPEC spectra of Class I disks from our sample -- although most of these spectra can be found on the {\tt SpExoDisks} Database \citep{SpExoDisks} -- for two reasons. In the literature, there has already been an extensive overview and analysis of rovibrational CO in these objects \citep{Herczeg2011}. 
Further information for these objects would require both an expanded sample and/or spectral imaging. Additionally, self-absorption makes it difficult to analyze line fluxes at the resolution of Keck-NIRSPEC \citep{Herczeg2011}. 

\section{Overview of Spectra}\label{sec:Analysis}

\begin{figure}
    \centering
    \includegraphics[width=\linewidth]{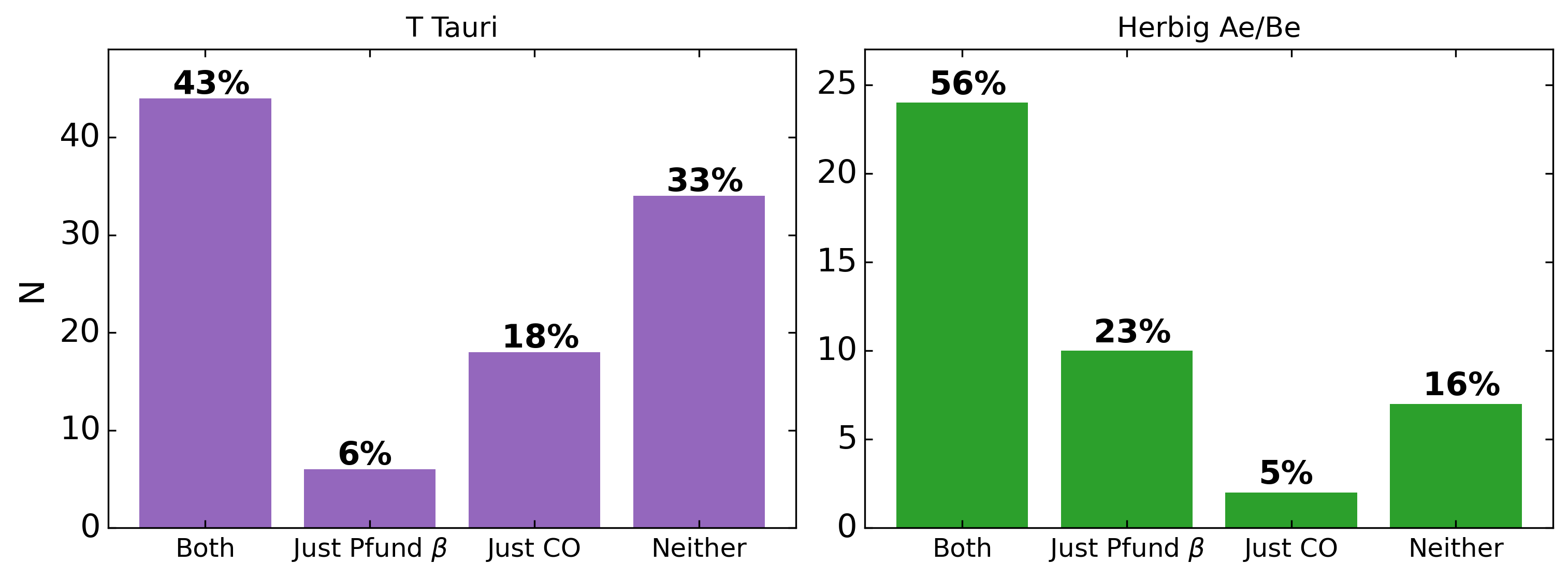}
    \vskip -0.1in
    \caption{Histogram for the differences in rovibrational CO and the H I Pfund $\beta$ (Pf$\beta$) emission detection rates separated by stellar subtype: TTS (left) and Herbigs (right). The percentage on top of each bar is relative to the total number of the stellar subtype.}
    \label{fig:PBRates}
\end{figure}

Rovibrational CO emission lines are quantified by the rotational quantum number, J. The P branch represents the transition from J-1 to J, while the R branch lines are from J+1 to J. For the purpose of this analysis, we define low J emission lines as the P(7) -- P(12) transitions and high J emission lines as the P(30) -- P(36) transitions. These regions were chosen based on our spectral coverage, and soundness of the telluric corrections. Included in the spectral range of most of our spectra is the mass accretion tracer H I Pfund $\beta$ (Pf$\beta$ 4.6538 $\mu$m). While we do not present a full analysis of this emission line in this work, we do present how often we detect Pf $\beta$ above a 3$\sigma$ threshold in our sample, separated by stellar subtype, in Figure~\ref{fig:PBRates}. 

\begin{figure}
    \centering
    \includegraphics[width=\linewidth]{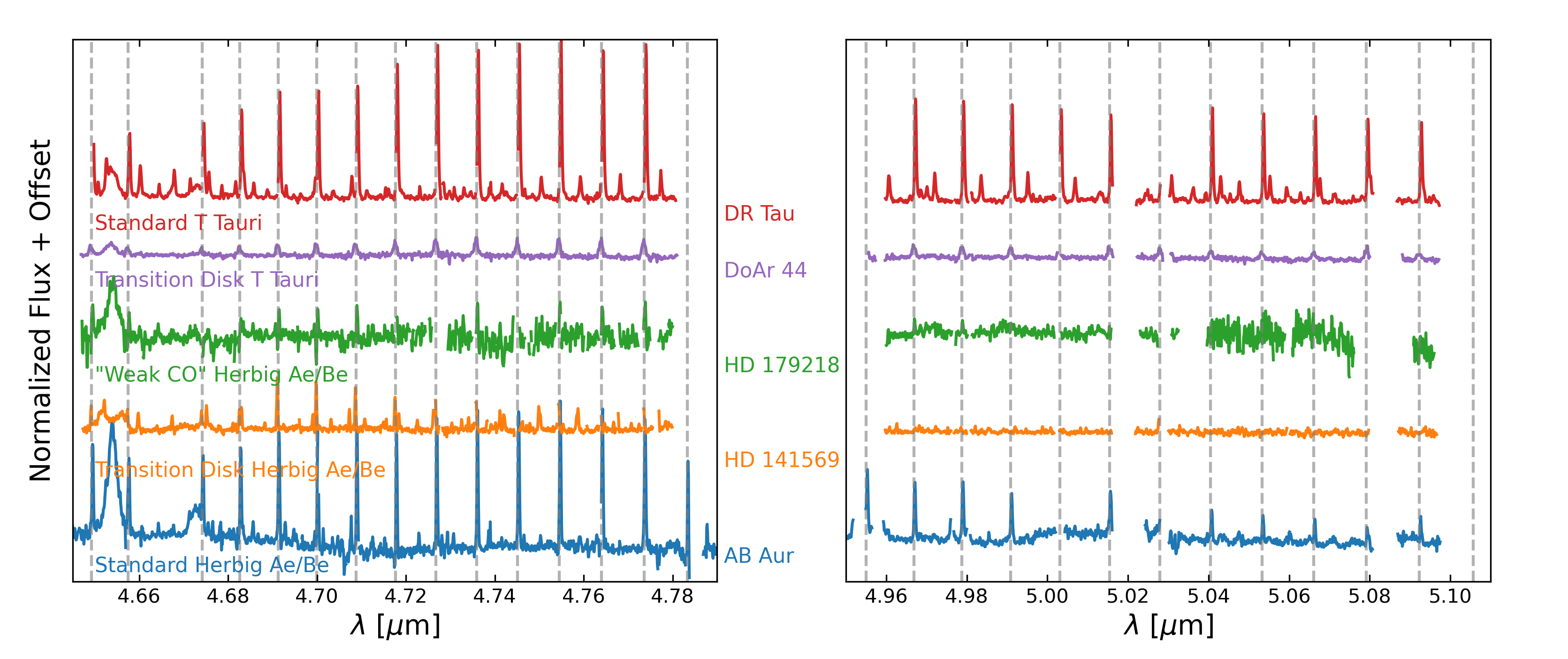}
    \vskip -0.1in
    \caption{Keck-NIRSPEC spectra that serve as examples as the 5 major types of rovibrational CO emission seen from Class II objects: strong CO CTTS (DR Tau), transition disk CTTS (DoAr 44), strong CO Herbig Ae/Be (AB Aur), weak CO Herbig Ae/Be (MWC 614), and transition disk Herbig Ae/Be (HD 141569). The location of all CO lines in the wavelength range are marked with a dashed gray line. In both of the strong CO examples, the $^{13}$CO and CO (2-1) transitions are also visible.}
    \label{fig:SubtypesSpec}
\end{figure}

Figure~\ref{fig:SubtypesSpec} highlights five examples from our sample that are representative of the most common classifications of objects with rovibrational CO in their spectrum. When there is CO emission in the standard TTSs, the full CO series is bright and apparent throughout the data. For the standard Herbig with CO emission, the entire CO series is also fully visible, but the entire series is less bright compared to the continuum, especially for the high J lines. Regardless of stellar mass, the transition disks tend to have lower line fluxes in the high J relative to their low J emission lines. The final example in Figure~\ref{fig:SubtypesSpec} highlights a Herbig with strong accretion, but ``weak'' CO emission lines -- these objects represents approximately half of Herbig subsample with accretion rates. The Pf $\beta$ emission that is linked to accretion is much brighter in this object than the CO, while in the other example objects the Pf $\beta$ emission is weaker or of equal brightness. 

Detections of rovibrational CO emission were determined by first checking every possible CO rovibrational emission line for the wavelength range of the spectra. If there were three or more emission features at CO rovibrational emission line ladder locations, the spectrum was considered to have CO rovibrational emission detected. 
Non-detections were followed up by creating stacked CO line profiles of every emission line location within the wavelength range. If there was a feature $>$3$\sigma$ in the stacked line profile above the noise level, then it was also considered a positive CO detection.
Both approaches take advantage of the large number of CO rovibrational ladder lines. We found an additional 16 objects with a rovibrational CO feature via stacking, but 11 of those objects had only an absorption feature rather than emission, which could additionally be due to telluric absorption corrections,  photospheric lines, or foreground interstellar material. 

All CO detections that display absorption rather than emission were initially classified separately. For objects that displayed CO with mixed absorption and emission features, the objects were included in this analysis as a CO emission detection. In some of these cases, where the emission component could be accurately approximated with a Gaussian, the sources are included in the proceeding analysis and results presented in this paper: there are 56 sources that have rovibrational CO absorption and of that subsample, 25 had emission components that could be a part of this further analysis. All other sources, 31 objects, with absorption are not included in the subsequent CO analysis.
For the most part, these objects had spectra that were classified with absorption on only a portion of their spectrum or a line shape that had distinguishable emission and absorption components: either a double-peaked line shape or a blue-shifted absorption component. For that subsample, the emission portions of the rovibrational CO lines was approximated via a Gaussian and the absorption portion of the signal was ignored for the line flux and slab modeling analysis. 

\begin{figure}
    \centering
    \includegraphics[width=0.45\linewidth]{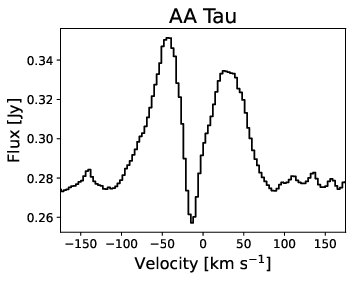}
    \caption{Selected $^{12}$CO emission lines are listed in Table~\ref{table:LineProfs}. The complete figure set (87 images) is available in the online journal.}
    \label{fig:LineProfs}
\end{figure}

Of the 183 objects in our sample, 96 objects in total have rovibrational CO emission lines detected using the criteria outlined above. Stacked line profiles were created via the combination of every non-blended $^{12}$CO emission detected in the spectrum for each object and are shown in Figure~\ref{fig:LineProfs} (the full figure with every object is available online). Due to the spectral resolution of NIRSPEC ($\sim$12 km s$^{-1}$), we present no analysis on the shapes or discussion on broad and narrow components and instead only utilize the full width at half maximum (FWHM) of the line shape, as all emission features in our sample had FWHM greater than the spectral resolution.

\subsection{CO Detection Statistics}\label{CODet}

\begin{figure}
    \centering
    \includegraphics[width=0.85\linewidth]{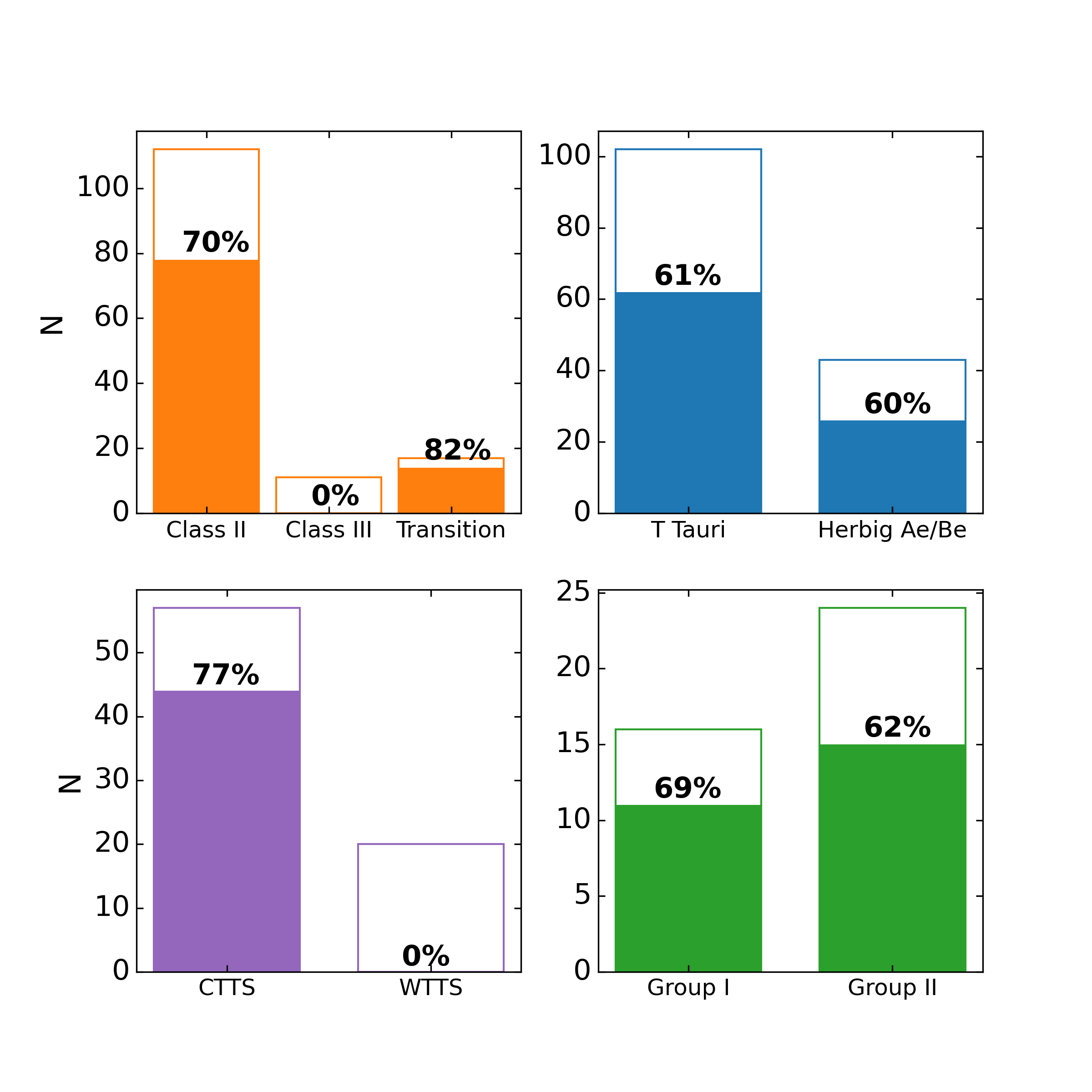}
    \caption{Histograms of the rovibrational CO emission detection rate in our sample as divided by (top left) SED classification, (top right) stellar subclass, (bottom left) TTS subtype, and (bottom right) Herbig Meeus group. The taller, empty bar represents the total number of each subtype within our sample, while the filled-in portion of the bar represents the portion of the subtype with rovibrational CO emission detected. Percentages marked note the percentage of CO detections for each grouping.}
    \label{fig:CORates}
\end{figure}

There are 78 confirmed class II objects that have rovibrational CO emission line detections, which is 82\% of the objects with CO emission lines detected; the other objects with detections are a combination of transition disks and objects without a confirmed SED classification. Within just Class II objects, TTS have a detection rate of 72\% and Herbigs 62\%. Figure~\ref{fig:CORates} shows our detection rates for four different classifications of objects in our sample: stellar SED classification (Class II versus Class III versus Transition Disk), stellar subclass (TTS versus Herbig), TTS subclasses (Classical versus Weak-lined), and Herbig Meeus Groups. There is not a significant difference in the presence of CO between the two main mass classifications of our sample: TTS (61\%) and Herbigs (60\%). Within the subtypes of each mass classification, there is some variation. Of the 17 transition disks in our sample, 13 of them have CO emission; which is greater than the overall Class II detection rate at 82\% -- most comparable to the CTTS detection rate of 77\%.

The presence of rovibrational CO emission separates the two TTS subclasses, with no WTTS having any CO emission detected while CTTS have the second highest rovibrational CO detection rate in our sample at 77\%. Since there are no WTTS with a rovibrational CO emission in our sample throughout the rest of this paper, outside of the detection statistics, if a TTS has a positive CO detection we classify it as a CTTS for the remainder of this work but not for the purpose of detection rates. The overall Class II detection rate is 70\% while zero Class IIIs have any rovibrational CO emission detected. 

Group I Herbigs have a slightly higher detection rate (69 $\pm$ 24\%), while Group II Herbigs display CO emission at slightly lower rates (58 $\pm$ 20 \%) -- although the difference could be accounted for within the uncertainties. 
Additionally, within both groups are subtypes; for an in-depth discussion on Herbig Meeus groups and rovibrational CO see \citet{Bosman2019} and \citet{Banzatti2018}. The $v2/v1$ ratio is sensitive to the structure of the inner disk of Herbigs (as discussed in \citet{Banzatti2018,Bosman2019}). Due to a lack of $^{12}$CO($2-1$) detections in the Herbigs (and the sample as a whole, most likely due to S/N), analysis dependent on the $v2/v1$ ratio was not attempted. 

The detection rates of this sample are comparable to those from \citet{Pontoppidan2010}, which reports detection rates of 60-80\%, dependent on spectral type. The detection rate of 61\% for Herbigs is slightly lower than prior reported detection rates ($\sim$80\%) but this is most likely due to higher noise levels in the Keck-NIRSPEC spectra. Compared to previous rovibrational CO emission surveys \citep[i.e.,][]{Pontoppidan2011,Brown2013,Banzatti2022} this sample is both larger ($\sim5\times$) and covers a more expansive stellar mass range. These previous samples have higher CO detection rates, but this is most likely due to the surveys being biased towards brighter disks.

There are 30 objects with accretion rates from the literature but no CO emission detection in our sample (see Table~\ref{table:Sample}). They are split evenly between Herbigs and TTS and cover a wide range of accretion rates. Their non-detections are for a few reasons: lower wavelength coverage, noisy spectra due to low continuum (either due to larger distance or dust depletion), older low-accreting TTS, and low Herbig line-to-continuum ratios (see Section~\ref{subsec:LF}). Of note, LkCa 15 has had a weak CO emission detected in the literature \citep{Najita2003,Salyk2009}, but due to noise in the spectrum, the stacked line profile did not qualify as emission according to the selection criterion utilized in this work.

\section{Results}\label{sec:Results}

\subsection{Line Fluxes}\label{subsec:LF}

We calculated line fluxes for each emission line detected for each object and a portion of these data are presented in Table~\ref{table:Linefluxes}, the full version is available in machine readable format. We also found average line fluxes for both the low J and high J spectral regions, by averaging the available transition line fluxes for each object. We use the low J and/or high J average line fluxes in place of individual transition line fluxes (i.e. P(6)) to account for some objects missing transition lines solely due to the reduction and correction process removing those parts of the spectra.  

\begin{figure}
    \centering
    \includegraphics[width=\linewidth]{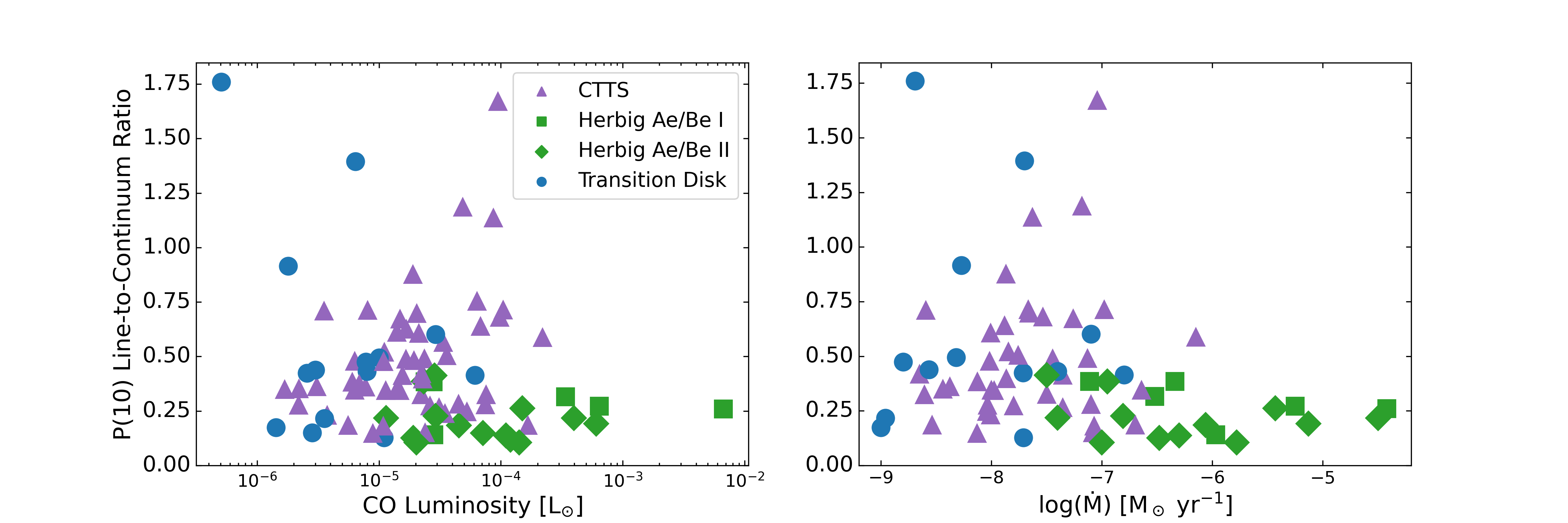}
    \caption{The P(10) CO line-to-continuum ratio versus the CO Luminosity (left) and the accretion rate (right).}
    \label{fig:LTCfig}
\end{figure}

Using the calculated line fluxes, we calculated two line-to-continuum (LTC) ratios: one as a representative low J ratio, P(10), and a high J, P(30). Figure~\ref{fig:LTCfig} illustrates trends in the P(10) LTC ratio versus the integrated line flux and accretion rate. There is a slight trend in the TTSs of increased LTC ratio with increased CO line flux, along with a tentative positive trend with accretion rate. The Herbigs show no trend with either CO line flux or accretion rate; they have overall lower LTC ratios and a smaller spread in their distribution than the TTSs. The transition disks appear to follow the same trend as the TTS, regardless of stellar mass, but tend to have lower LTC ratios when considering the P(30) LTC. This is consistent with a general trend in transition disks of having fewer detectable high J transitions than their Class II counterparts. 

\begin{figure}
    \centering
    \includegraphics[width=0.85\linewidth]{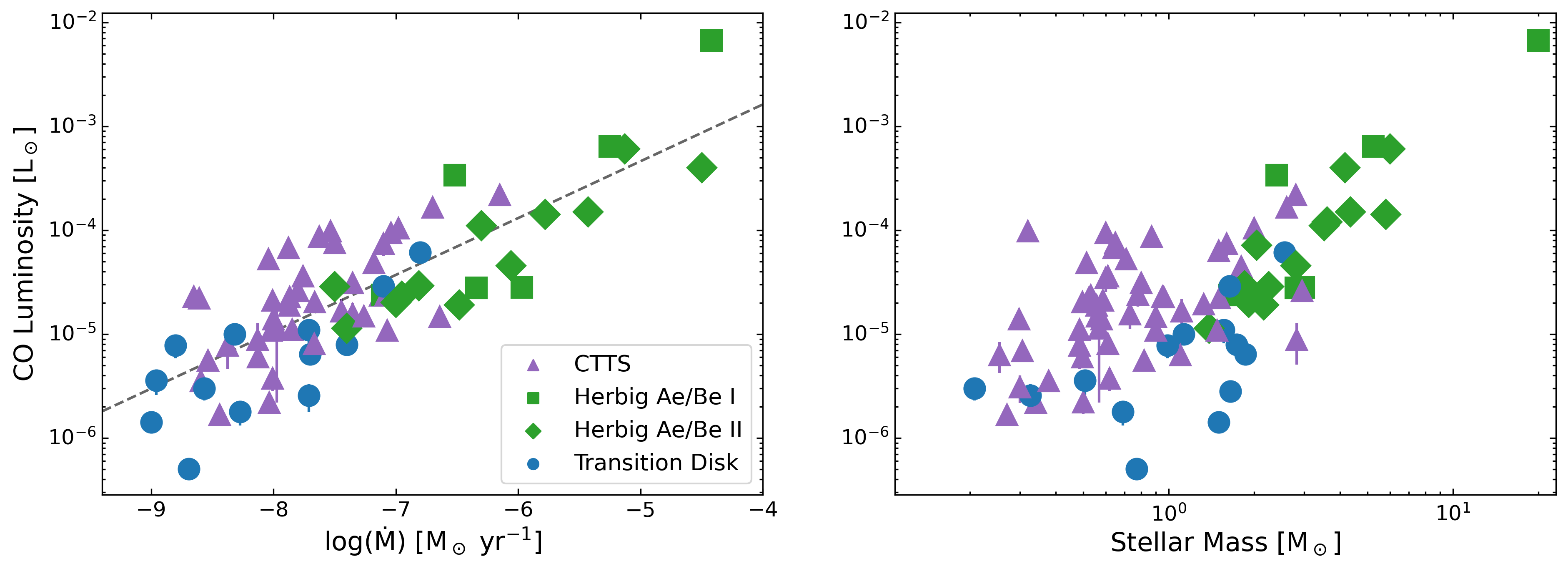}
    \caption{Left: The CO Luminosity versus the stellar accretion rates (Table~\ref{table:Sample}).  Object subclasses are marked by the symbols denoted in the top left corner. The gray dashed line is the linear fit for the full sample in log-log space. Right: The CO Luminosity versus the stellar mass (Table~\ref{table:Sample}). Object symbols are the same as the left panel.
    }
    \label{fig:lumaccretion}
\end{figure}

The left panel of Figure~\ref{fig:lumaccretion} shows the correlation between CO luminosity and accretion rate. We chose to investigate accretion rate rather than accretion luminosity to take advantage of the accretion rates presented by M23 and \citet{Betti2023}.
We find there is a positive correlation between CO luminosity and accretion rate for all sources, regardless of stellar properties. The trend has a Spearman coefficient of 0.73, which is indicative of a strong positive correlation. Using linear regression, we find a $p$-value of 3.49$\times 10^{-16}$ for a positive correlation and a fit of:
\begin{equation}
    \log L_{CO} = (0.54\pm0.05)\log \dot{M} - (0.61\pm0.38)
\end{equation}

where $L_{CO}$ is the CO luminosity in L$_\odot$ and $\dot{M}$ is the accretion rate M$_\odot$ yr$^{-1}$.

\subsection{Rotation Diagrams}\label{subsec:RotDia}

Rotation diagrams are plots that compare the line strength versus the excitation energy of a rotational series. There are no assumptions made in the plotting of a rotation diagram, but their interpretation generally involves making different degrees of assumptions. The x-axis of a rotation diagram is the excitation energy (K) of each line, while the y axis is the natural log of the line flux normalized by the line properties, representative of the level population. Rotation diagrams arose primarily for the study of optically thin tracers; in the optically thin limit, if the gas is at a single temperature, the rotation diagram becomes a straight line, with a slope given by $-1/T$, where $T$ is the gas temperature. In the optically thick limit, the lower excitation levels (i.e. the highly populated ones) of the rotation diagram begin to curve, due to their saturated flux and the increase of the level degeneracy, $g$, at low J. The curvature in the lower excitation levels can also be caused by non-LTE effects such as IR and UV fluorescence \citep{BlakeBoogert2004,Thi2013}. Non-LTE effects can also affect the line strengths in the higher energy levels due to the higher critical densities and possible subthermal population levels. 

\begin{figure}
    \centering
    \includegraphics[width=\linewidth]{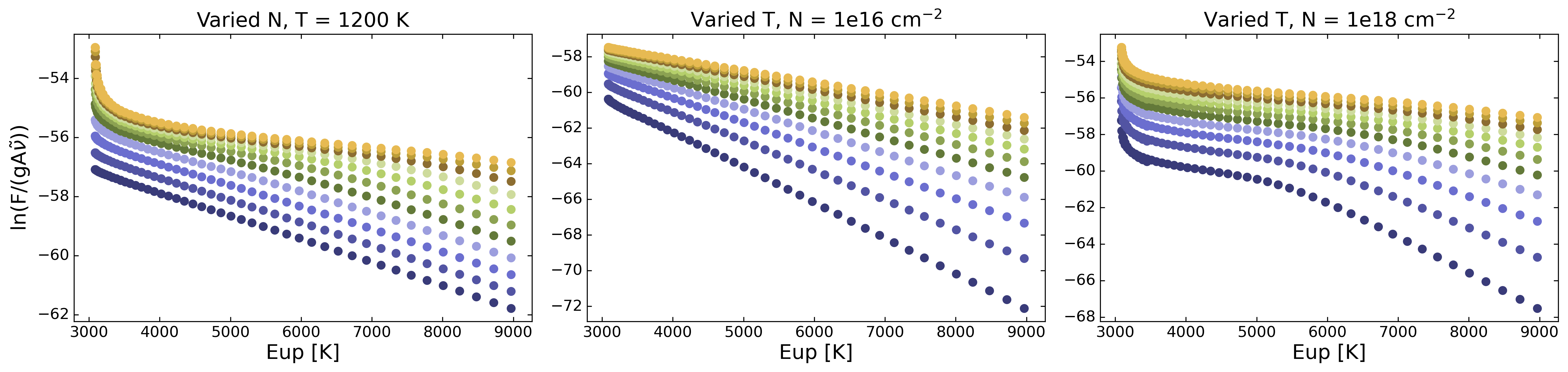}
    \caption{Generated slab models of rovibrational CO in a disk with different temperatures and column densities. In the left panel, temperature is held constant (at 1200 K), while only column density is varied from 10$^{16.25}$ $-$ 10$^{18.75}$ cm$^{-2}$ with steps of $0.25$ in log space. The lower rotation curves (purple) represent lower column density, representing optically thin disks, while the higher rotation diagram curves (yellow) have higher column densities and illustrate how optical thickness effects the shape of a rotation diagram. In the center and right panels, column density is held constant (at 10$^{16}$ cm$^{-2}$ and 10$^{18}$ cm$^{-2}$ respectively) while temperature is varied from 500 $-$ 1500 K with steps of 100 K. Lower temperatures (purples), have steeper slopes, while higher temperatures (yellows) are flatter.}
    \label{fig:RotDias}
\end{figure}

Rotation diagrams from generated slab models are shown in Figure~\ref{fig:RotDias}. The effect of changing column density on the rotation diagram is shown in the left panel of Figure~\ref{fig:RotDias}. Optical depth effects are dependent on the column density of the gas and affect the shape of the rotation diagram. As the column density, and therefore each level population increases, the optical depth approaches 1 and the line fluxes plateau; this causes the rotation diagram to appear curved at the lower energy levels, which are more highly populated and therefore experience optical depth effects first. The center and right panel of Figure~\ref{fig:RotDias} illustrates the effect of the gas temperature on the slope of the rotation diagram for both optically thin (center) and optically thick scenarios (right). The temperature of the gas changes how the energy levels of the gas are populated; this changes the slope of the rotation diagram to be flatter with higher temperature due to the higher fluxes in the higher excitation levels. In the optically thick scenario, the changing temperatures still affect the slope of the rotation diagram but only in the higher energy levels.

\begin{figure}
    \centering
    \includegraphics[width=\linewidth]{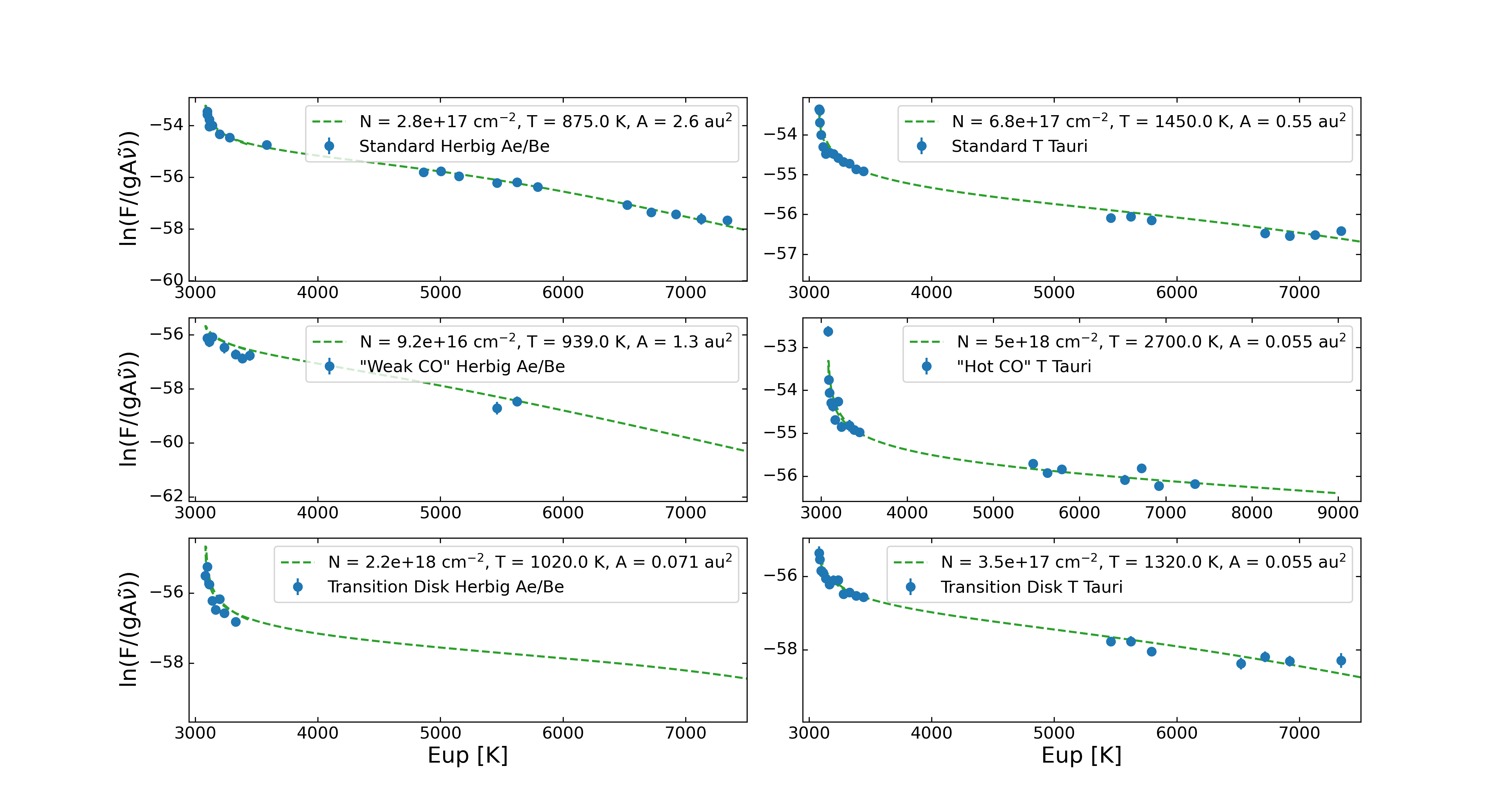}
    \vskip -0.1in
    \caption{Example rotation diagrams for the six objects presented in Figure~\ref{fig:SubtypesSpec}: strong CO CTTS (DR Tau), hot CO TTS (CI Tau), transition disk TTS (DoAr 44), strong CO Herbig Ae/Be (AB Aur), weak CO Herbig Ae/Be (HD 179218), and transition disk Herbig Ae/Be (HD 141569). The data are plotted in blue circles and a slab model rotation diagram generated from our best fit parameters is plotted as a green dashed line.}
    \label{fig:SubtypesRotDias}
\end{figure}

Rotation diagrams that exemplify the sample are illustrated in Figure~\ref{fig:SubtypesRotDias}. If the slope of the higher energy levels are interpreted as solely due to temperature, the TTS disks have the highest temperature of the 5 example types. The Herbig, regardless of the presence of an inner gap, shows less curvature in the low energy levels, which implies either (or both) lower column densities and less non-LTE effects. The transition disks also have less curved rotation diagrams, which is again associated with lower column densities and/or less non-LTE effects. 

\subsection{Slabspec Modeling}\label{subsec:SlabSpec}
The limitations of rotation diagram interpretations can be partially bypassed by using a slab model that allows for optical depths $>$1. These models treat the gas as a slab of constant column density and temperature. In this work, we utilize the slab modeling codes ``{\tt slabspec}'' and ``{\tt slab\_fitter}'', included in the {\tt spectools\_ir} package \citep{spectools}.  

The three free parameters of the slab model are the temperature, the solid angle, and the column density. The best slab model for each object is generated using MCMC retrievals. The priors for these retrievals were informed by each object's rotation diagram and distance. If the rotation diagram showed curvature in the low energy levels an optically thick prior was used and if not, an optically thin prior was selected. The range for the solid angle is determined by the distance to the object and the range of radii that are valid for CO emitting regions based on previous work \citep[$0.01 - 10$ AU,][]{Salyk2011b}. The maximum temperature used for most objects is the dust sublimation temperature of 1750 K, but for some TTS with very flat rotation diagrams, we raised the upper limit to 3000 K. This temperature is an estimate of the thermal dissociation temperature for CO based on when CO bands are seen in the spectra of stars \citep[this happens as high as 5000 K but is more commonly seen around 3500-3000 K, ][]{Heras2002}.

For the 96 objects with detectable CO,  we were able to perform slab modeling and retrieval on 67 (70\%) of them. 
The other 22 objects with CO had enough emission lines to be sure of the presence of CO, but had less than 5 usable emission lines. Often, this was due to exclusion of low telluric transmission regions leading to crucial parts of the emission line to be missing from the spectra, CO absorption not leaving enough recoverable information about the emission line shape, and/or low signal-to-noise ratios (S/N) suppressing most of the CO ladder below the noise. The retrieval cannot confidently output results for these objects, as the entire range of parameters can fit the data.

The retrieved properties (temperature, column density, and emitting area) for the successfully modeled objects are presented in Table~\ref{table:RetrievedProps}. Examples of model rotation diagrams generated from slab models of the retrieval best fit parameters are plotted in Figure~\ref{fig:SubtypesRotDias}. An additional subtype for the sample has been added for this portion: the ``hot CO'' TTS which are the TTS that required the very high upper limits on temperature mentioned previously. The area output of the slab modeling is in solid angle, which has been converted to emitting areas using the distances listed in Table~\ref{table:Sample}. We also calculated a deprojected emitting area using the sub-mm inclinations of the disks, or 45$^\circ$ if there was none in the literature. When the area is referred to, that is the area without accounting for inclination, while the deprojected area is referred to thusly and was used to calculate the retrieved radii (R$_{ret}$).

Overall, the slab models generate spectra that match the observed rotation diagrams of the objects well. The retrieved values are better constrained when there are more lines detected, that include a broad range of excitation energies. The higher excitation level transitions are more sensitive to both the smaller and hotter emitting regions of CO, which in some cases biases the temperatures to warmer temperatures and smaller retrieved areas. This effect will only highly affect the retrieved properties when the broad component contributes significantly more flux than the narrow component. We find in our sample that the narrow component tends to dominate the flux significantly enough in the lower J, that the retrieved properties remain more statistically similar to the narrow component. A subset of objects with the highest temperatures present in Table~\ref{table:RetrievedProps} are potentially a byproduct of this effect, and the retrieved properties are representative of the smallest, hottest region of CO in the inner disk of these objects.

\begin{figure}
    \centering
    \includegraphics[width=0.32\linewidth]{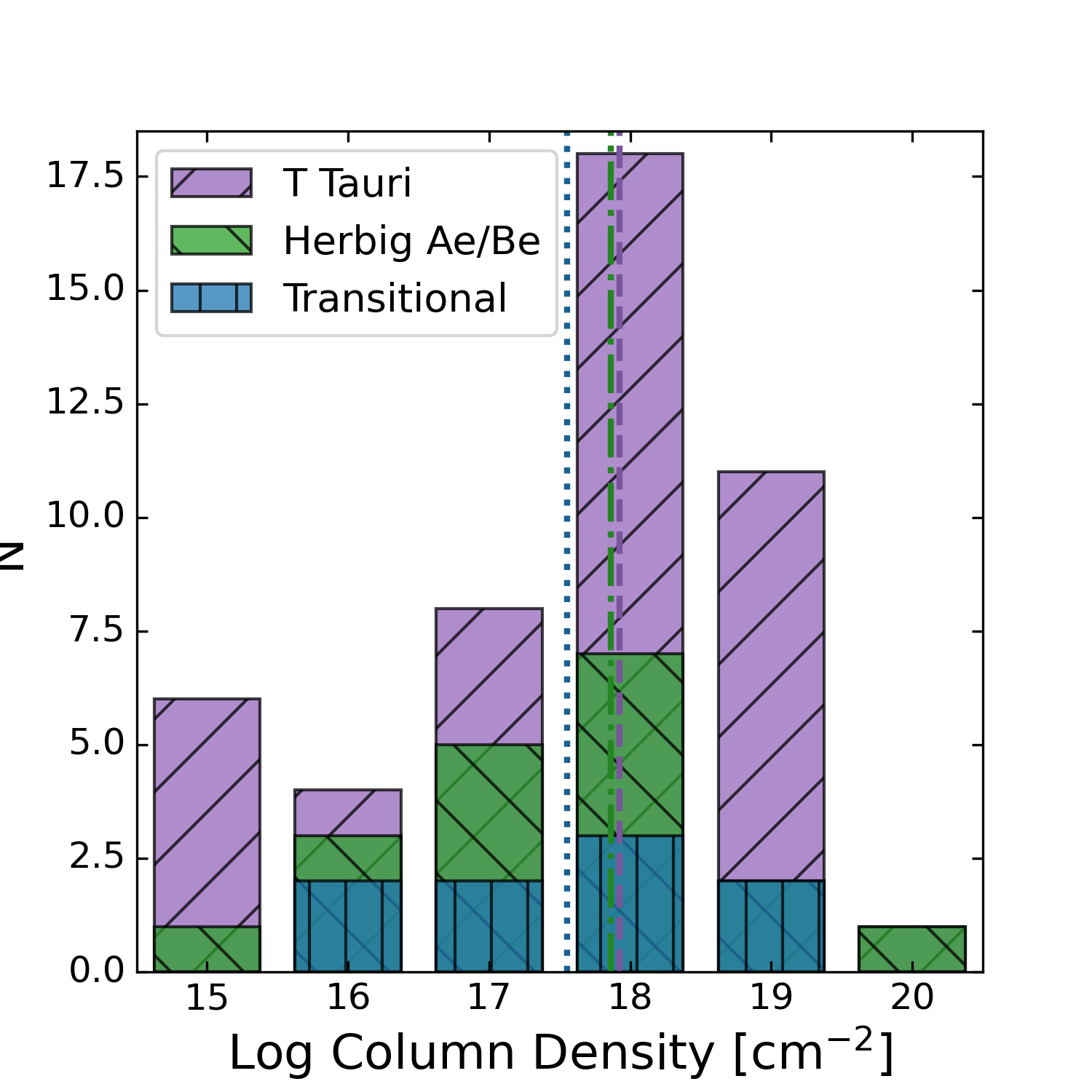}
    \includegraphics[width=0.32\linewidth]{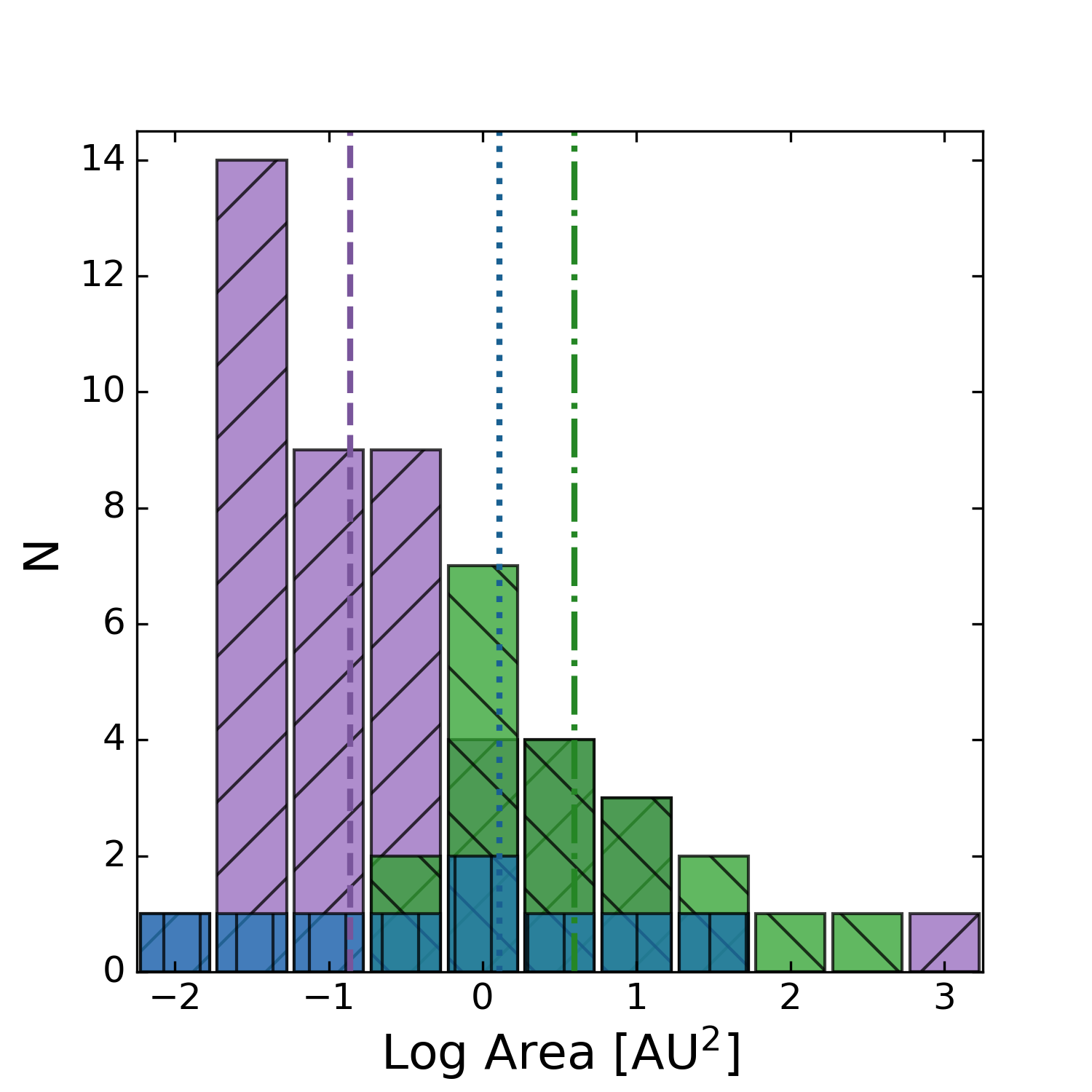}
    \includegraphics[width=0.32\linewidth]{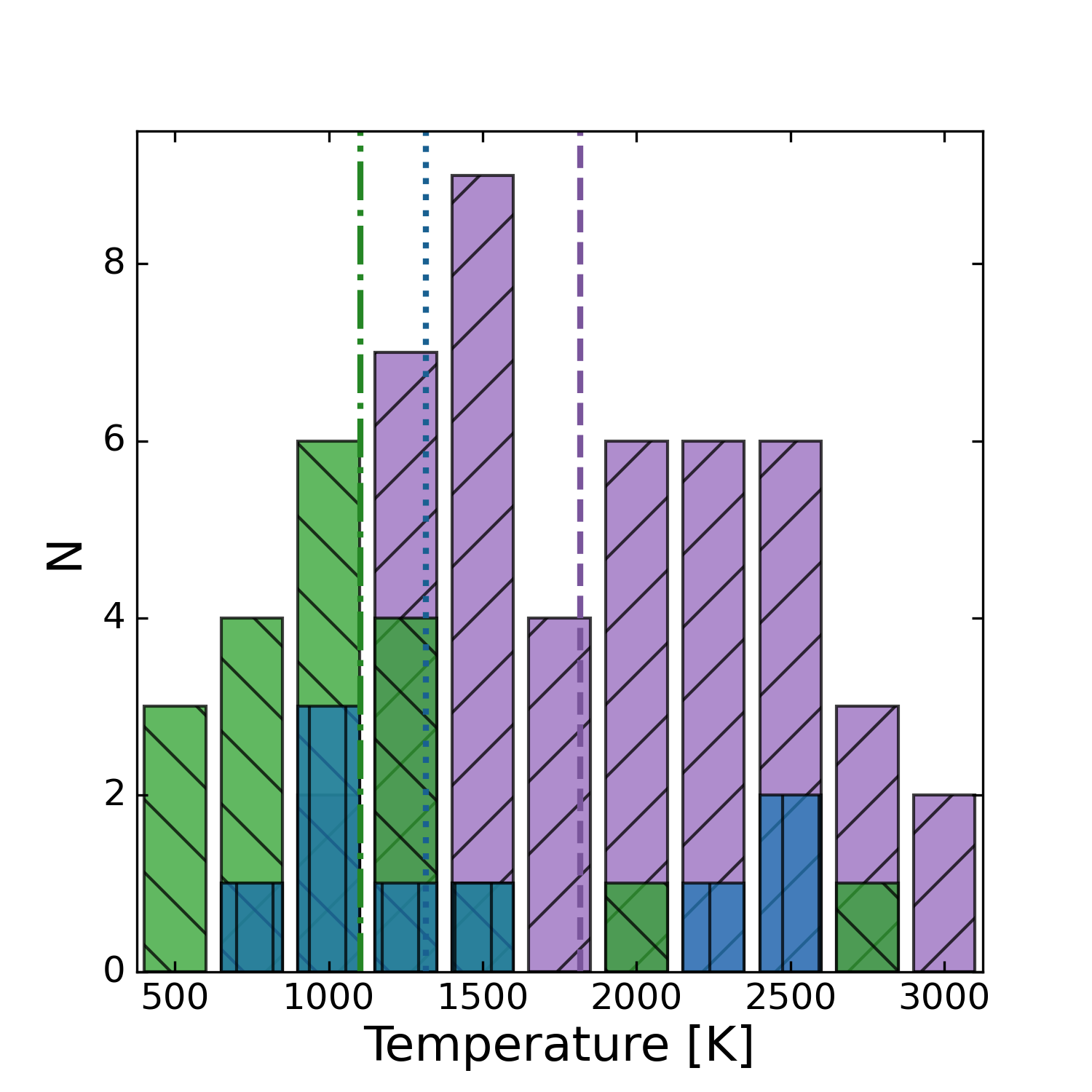}
    \caption{Histograms for the retrieved properties for the entire sample separated by disk subtype: TTS (purple, right hatch, dashed line), Herbig (green, left hatch, dash-dotted line), and Transition (blue, vertical hatch, dotted line). From left to right, the retrieved properties are column density, emitting area, and temperature. The vertical lines are representative of the medians for each subsample, colored to match their respective subtype.}
    \label{fig:RetResults}
\end{figure}

Histograms of the overall behaviors of the three retrieved properties are presented in Figure~\ref{fig:RetResults}. There is no major variation in the distribution of column density between Herbigs and TTSs, the most common retrieved column density being around 10$^{18}$ cm$^{-2}$. The TTS distribution does extend to lower column densities, potentially because of higher line-to-continuum ratios in TTS making the retrieval more sensitive to lower column densities. There is a difference in the distribution of properties between the Herbigs and TTSs for both the emitting area and temperature: Herbigs have larger emitting areas and TTSs have higher gas temperatures. Seven of the thirteen transition disks with CO emission had successful retrievals and while there is not enough of a sample to declare their behavior with certainty, they appear to align more with the Herbigs distribution for all three properties. 

In some optically thin cases, the retrieval is able to solve for the temperature well, but not the other two parameters due to the degenerate nature of the column density and emitting area. This is also the case if the low J-levels are not detected -- due to possible absorption, telluric corrections, or missing data -- meaning there were none or few low excitation transitions to assist the fitting. These cases are denoted in Table~\ref{table:RetrievedProps} and will be marked in all plots with retrievals by a plus sign.

\subsection{Line Widths}\label{subsec:Linewidths}
\begin{figure}
    \centering
    \includegraphics[width=0.95\linewidth]{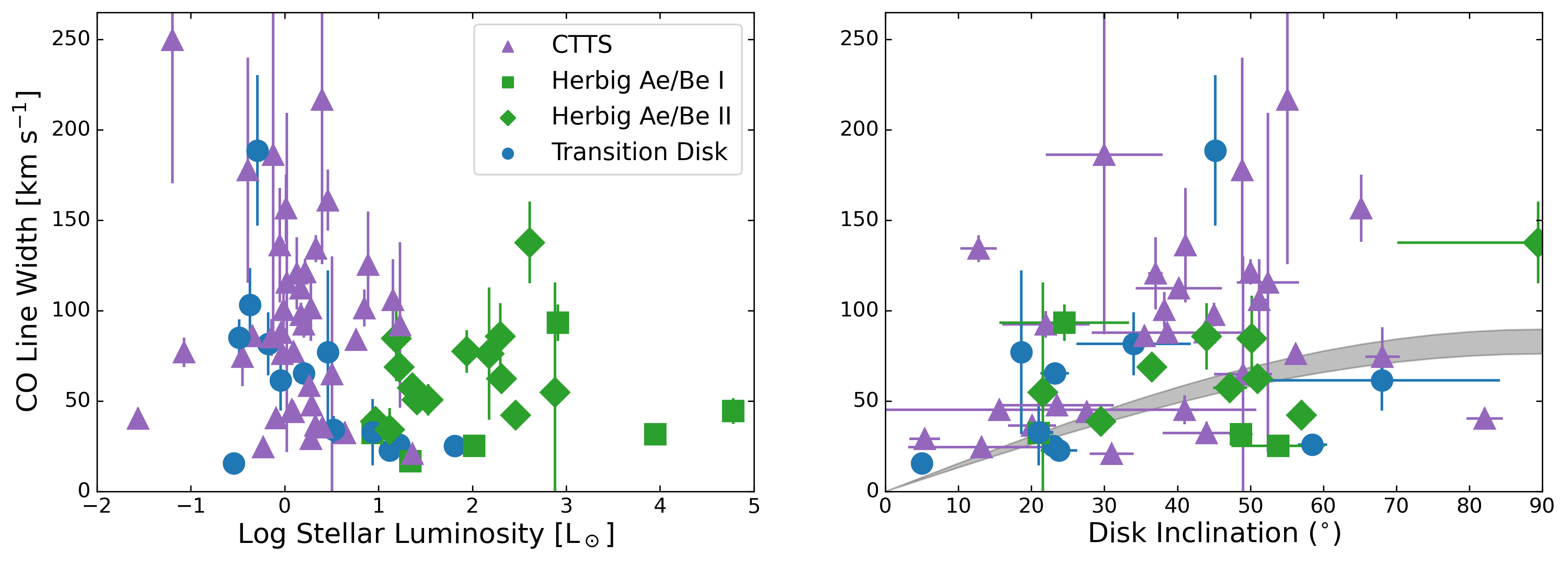}
    \caption{Left: The average CO emission line width versus stellar luminosity with the stellar subtypes marked as in legend. Right: The same CO line width versus mm-dust disk inclination. The gray-shaded region represents the range of predicted broadening due to Keplerian rotation at the sublimation radius around a K7 TTS to a A5 Herbig.}
    \label{fig:linewidth}
\end{figure}

The line width of the CO emission is a tracer of the velocity of the gas: in the simplest scenario, broadening of the line is solely due to the Keplerian velocity of the gas at the emitting radii. Average line widths for each object were calculated by measuring the FWHM of each non-blended CO emission line detected and then averaging these for a representative value. The errors associated with each line width is the standard deviation for the selected lines. The lines used to determine this value, as well as the average line width, for each object are listed in Table~\ref{table:LineProfs}.

Figure~\ref{fig:linewidth} illustrates the correlation of line width versus stellar luminosity, showing an overall pattern where CTTS tend to have broader CO features than their Herbig counterparts, but the dispersion in line widths observed in CTTS is very large. Line broadening may be expected to anti-correlate with stellar luminosity, because the temperature of the gas that excites CO extends to farther disk radii as the stellar luminosity increases, decreasing the line widths. 

Previous studies of CO line kinematics have observed that the line widths are linked to both the inclination of the disk and also the presence of inner cavities \citep{BlakeBoogert2004,Salyk2011b,Banzatti2022}. This trend holds true overall for this sample, the line width trending positively with disk inclination (Figure~\ref{fig:linewidth}, Right). The dispersion in the trend is most likely due to the emitting regions of the CO emission also having an effect on the outer radii of the disks -- inner gas depletion removes CO emission at high velocities and results in narrower line profiles. 

\subsection{Emitting Radii}\label{subsec:EmitRad}

Three different radii were calculated for the sample: an inner radius, R$_{\rm CO}$, a retrieved radius, R$_{ret}$, and an experimental outer radius, R$_{out}$. We calculated the inner radius via using the 1.7$\times$HWHM \citep{Salyk2011b} of the stacked line profile as a proxy for Keplerian broadening ($R_{CO} = GM(\frac{\sin{i}}{1.7\times HWHM})^2$). While \citet{Banzatti2022} suggests that the inner radius is better estimated by the half-width-at-10\%, we use R$_{\rm CO}$ as our estimate of the inner radius estimate following \citet{Salyk2011b}. The retrieved radius was calculated from the retrieved solid angles assuming the emitting area is a circle projected by the disk inclination $i$ ($R_{ret} = \sqrt{\frac{A}{\pi\cos{i}}}$). For cases where the disks had no available millimeter-wave outer disk inclination measurements, an inclination of 45 degrees was assumed. Inclinations are listed in Table~\ref{table:Sample}. 
Following \citet{Salyk2011b}, the inner radius of the disk correlates with the area found from the rotation diagram by estimating the emitting area as a ring ($A = \pi(R_{out}^2 - R_{in}^2)$). Therefore, we calculated the experimental outer radius (R$_{out}$) with the assumption that the retrieved emitting area is a ring with an inner radius of R$_{\rm CO}$. The inner radius, R$_{CO}$, along with the FWHM of the stacked line profiles are listed in Table~\ref{table:LineProfs} and the outer and retrieved radii are listed in Table~\ref{table:RetrievedProps}. All radii and the resulting geometry of the emission will be discussed further in Sec.~\ref{subsec:EmitArea}.

\section{Discussion}\label{sec:Discussion}

\subsection{{\rm CO} as an Inner Disk Tracer}\label{subsec:InnerDiskTracers}

For stars with disks, CO is potentially one of the most robust tracers of inner disk presence. Besides rovibrational CO lines there are a few other commonly used inner disk tracers, with each probing different mechanisms tied to the inner disk: rovibrational CO directly traces the bulk gas content, while other tracers are more indirect. Accretion is a robust indicator of the presence of an inner disk and is observed in disks via a multitude of lines, like Pf $\beta$ which is in our wavelength range. While for the TTSs, rovibrational CO emission lines are more sensitive to the presence of a gaseous inner disk, as illustrated in Figure~\ref{fig:PBRates}, Pf $\beta$ appears to be a better tracer for Herbigs. The presence of dust (i.e. via identification of an infrared excess) does not guarantee the presence of a detectable rovibrational CO emission line, but the lack of a dust disk does provide insight into which sources will not have any detectable CO. 

Based on the detection rates of this sample, CO is one of the most effective tracers for the gaseous inner disks of CTTS (see Figures~\ref{fig:PBRates} and~\ref{fig:CORates}). 
While accretion tracers tend to be a good indicator of disk presence, they are rarely found without CO rovibrational emission in CTTS - 88\% of TTS with Pf $\beta$ also have rovibrational CO emission lines. 
The 6 remaining TTS disks with just Pf $\beta$ all have relatively low continuum fluxes and higher SNRs, but have moderate to high accretion rates which means the broader and brighter Pf $\beta$ feature is still visible.

For Herbigs, the presence of rovibrational CO emission lines are slightly less effective at probing the inner disk (only 59\% of the Herbigs in our sample had a CO detection). Within the subgroups, 12 of the 17 ($\sim$69\%) confirmed Meeus Group I Herbigs had a CO detection while 14 of the 25 ($\sim$58\%) confirmed Meeus Group II Herbigs had a CO detection. This stands in contrast to the traditional morphological distinction between the two groups - where Group II Herbigs are thought to have either no inner cavity or only a small one, while Group I Herbigs have a large gap in their disk. 

\subsection{Accretion Heating of CO}

\citet{Najita2003} found that the strength of CO emission correlates with indicators for disk accretion (both the mass accretion rate and the K-L color). Figure~\ref{fig:lumaccretion} illustrates that all objects with CO emission have line fluxes that positively correlate with accretion rate. Dispersion from the trend is most likely due to variability as young TTS and Herbigs experience variability not only in stellar luminosity but also in accretion rate \citep{Costigan2012}. While accretion rate and stellar mass also correlate \citep{Alcala2017}, the stellar mass trend is more moderate with a Spearman coefficient of 0.58 and linear regression in log space yields:
\begin{equation}
    \log L_{CO} = (1.18\pm0.14)\log M_* - (4.72\pm0.06)
\end{equation} 
This slope is similar to the trend reported in \citet{Alcala2017} between $\dot{M}$ and $M_*$ (slope $=$ 1.3$\pm$0.24), which potentially again links the accretion rate and the CO luminosity.

\begin{figure}
    \centering
    \includegraphics[width=0.45\linewidth]{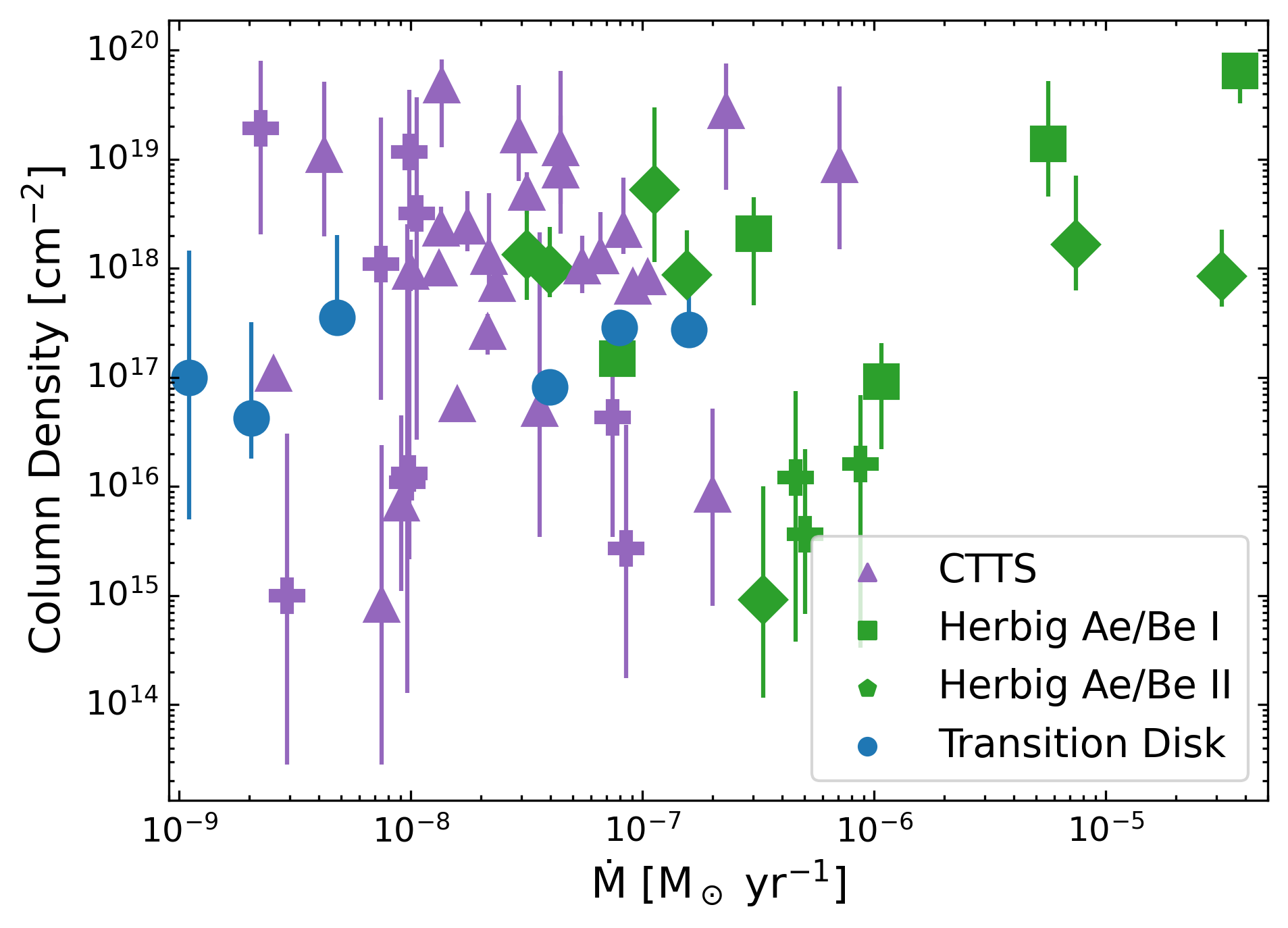}
    \includegraphics[width=0.45\linewidth]{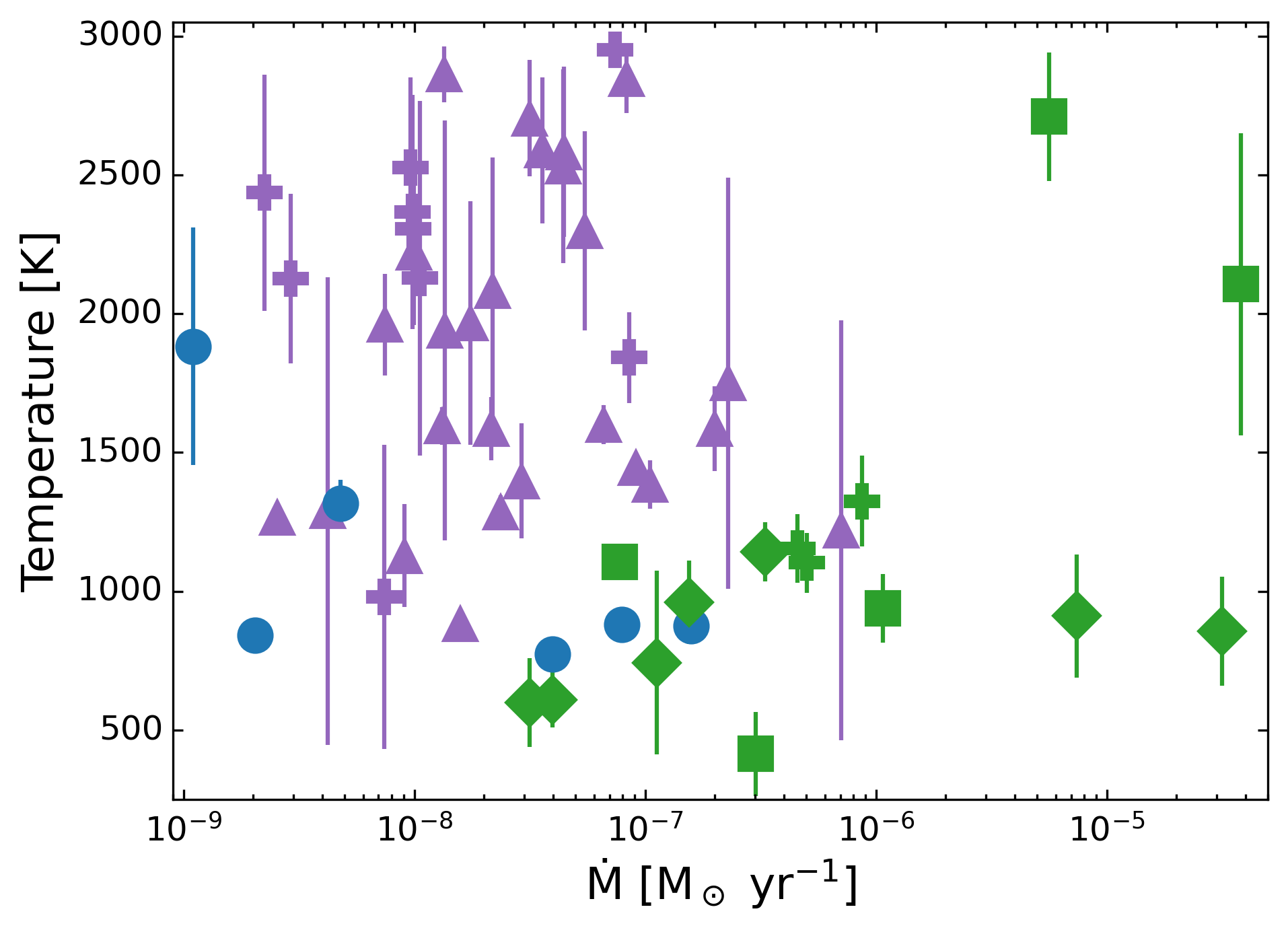}
    \caption{Left: The retrieved column densities versus the accretion rate for all sources with retrievals and accretion rates. The stellar and disk subtypes are as marked in the legend. Right: The retrieved temperatures versus the accretion rate for the sample. Poor retrievals are marked with the plus sign, as defined in Section~\ref{subsec:SlabSpec}.}
    \label{fig:RetAcc}
\end{figure}

There does not appear to be a link between retrieved column density and accretion rate (Left, Fig~\ref{fig:RetAcc}), but
the retrieved temperature does tentatively correlate with an increased accretion rate in CTTS (Right, Fig~\ref{fig:RetAcc}). The latter correlation has a Spearman's value of 0.19, which is a weak positive correlation, and a p-value of 0.26. This correlation could be due to the retrieved temperatures being inflated in CTTS due to the high excitation transition emissions probing the innermost, hottest regions of the disk, i.e. where the material is actively accreting onto the star. CO temperature could be linked to the accretion rate via viscous processes due to increased accretion rates heating the disk. 
Unlike the correlation between accretion rate and CO luminosity, the temperature of Herbigs trends separately from the CTTS. Almost all Herbigs have temperatures between 450 and 1500 K regardless of accretion rate, with the exception of two high accretors with high temperatures. 

The difference in temperature versus accretion rate trends in Herbigs versus TTS is possibly unrelated to accretion rates and instead to stellar luminosity -- Herbigs have higher stellar luminosities and therefore the additional luminosities provided by accretion are relatively smaller and therefore less impactful on the temperature of CO emission reservoir. As noted, Herbigs overall have lower LTC ratios that are unrelated to either stellar luminosity or accretion rate, while CTTS LTC ratios seem to positively trend with both.

\begin{figure}
    \centering
    \includegraphics[width=0.33\linewidth]{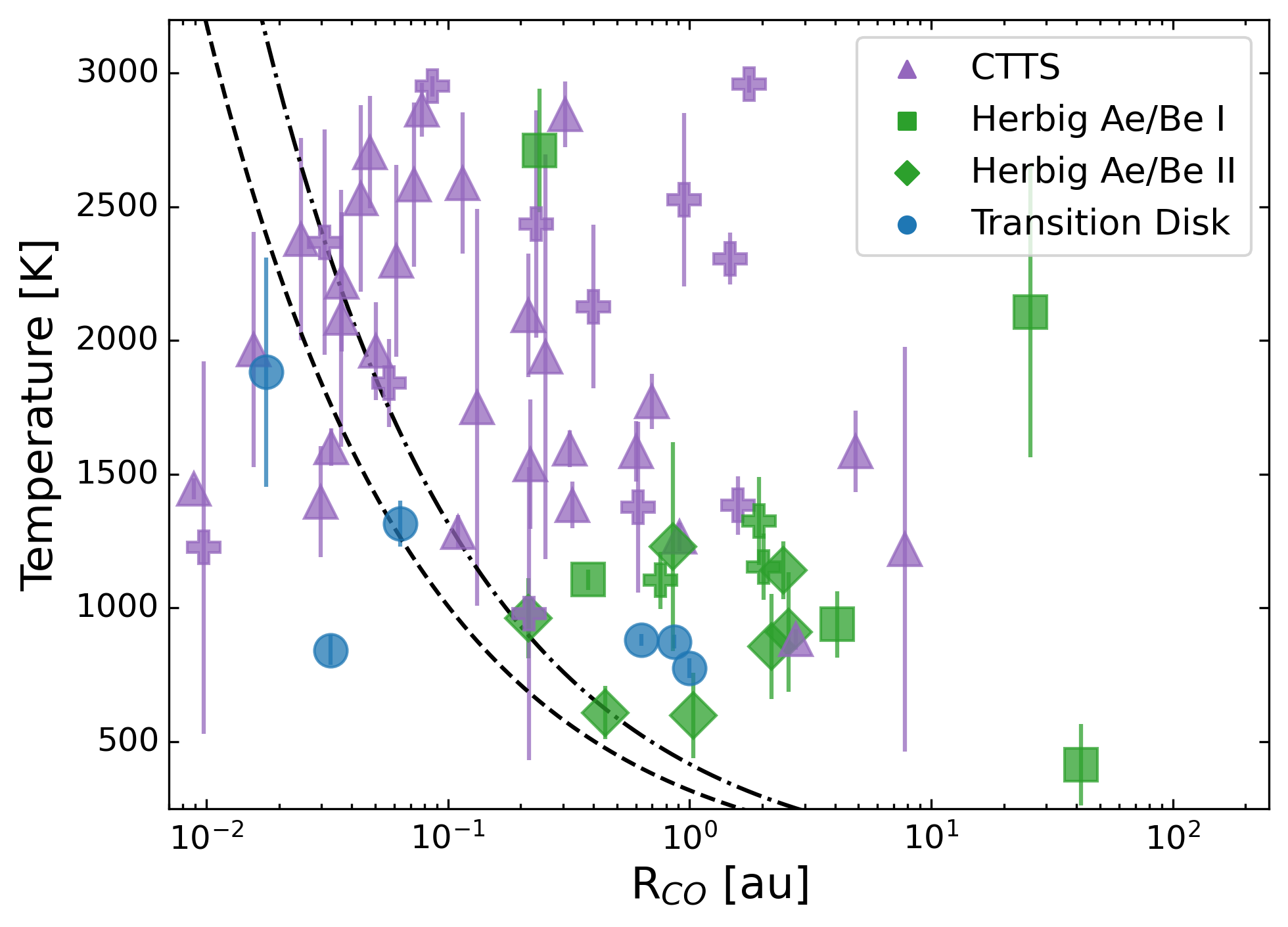}
    \includegraphics[width=0.32\linewidth]{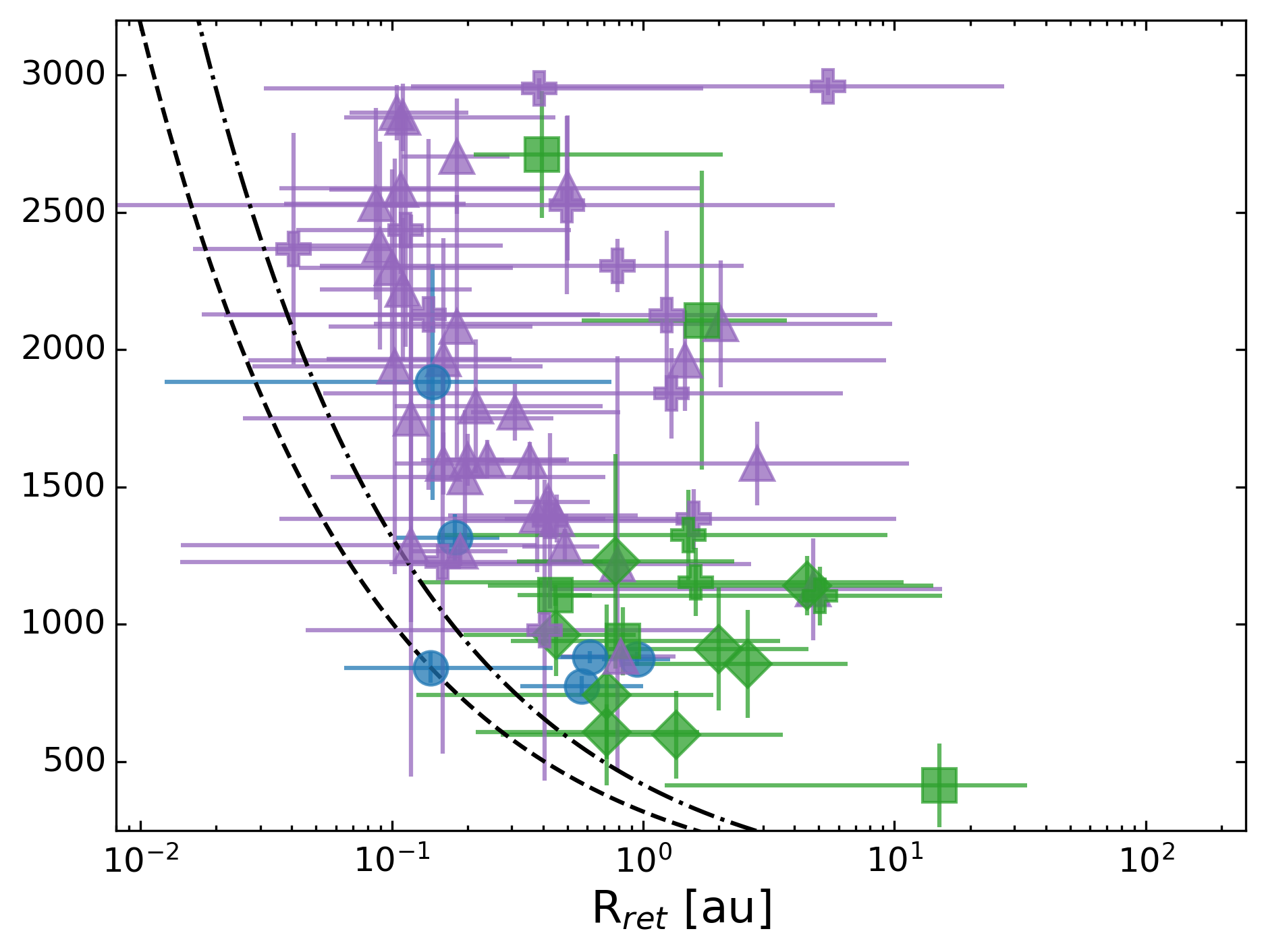}
    \includegraphics[width=0.32\linewidth]{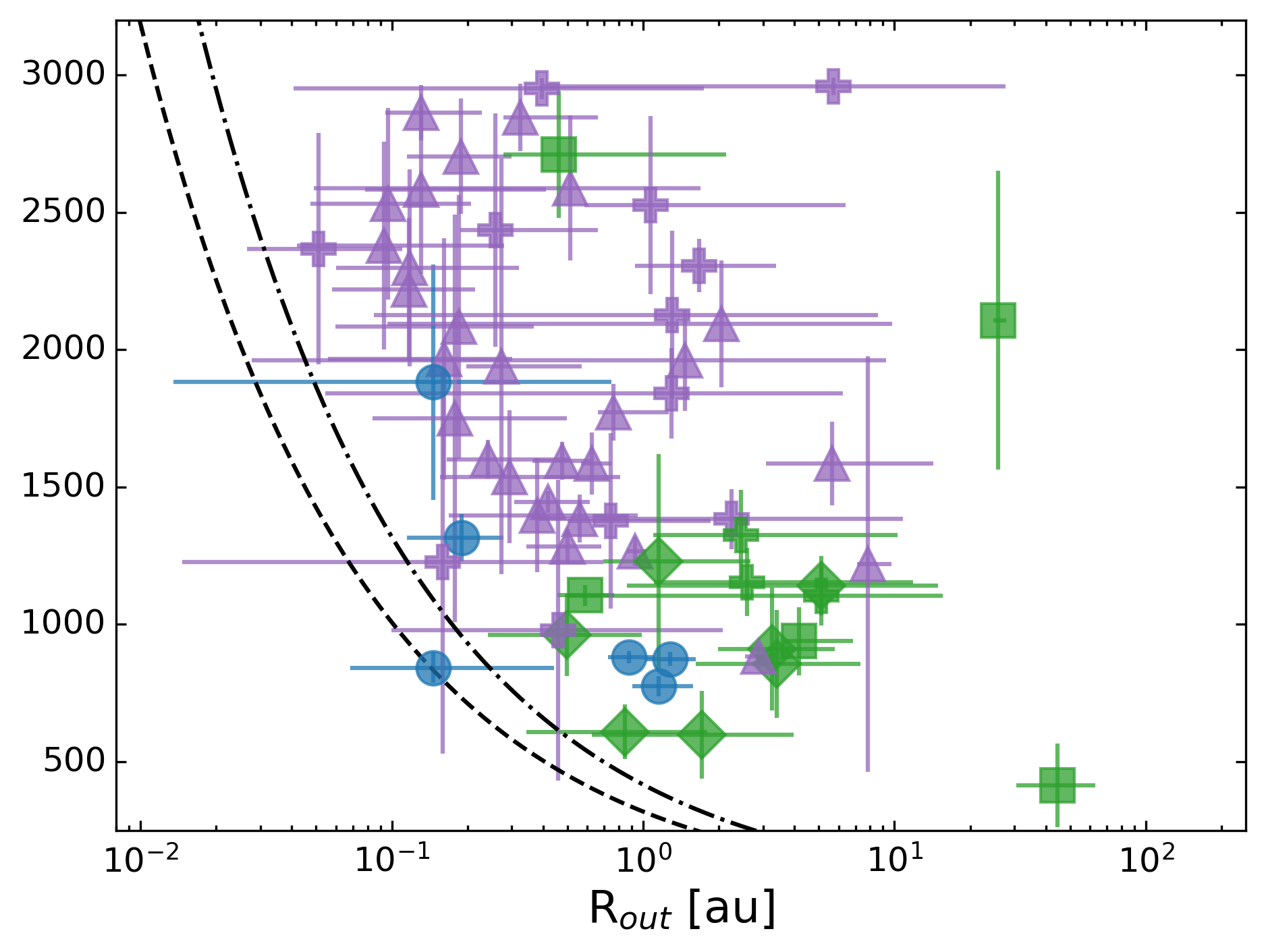}
    \caption{Left: Retrieved temperature versus R$_{CO}$. Middle: Retrieved temperature versus R$_{ret}$. Right: Retrieved temperature versus experimental outer radius, R$_{out}$. The dashed lines are the surface dust temperatures for large and small grains calculated via the prescriptions from \citet{Dullemond2001}. Poor retrievals are marked with the plus sign, as defined in Section~\ref{subsec:SlabSpec}.}
    \label{fig:TempRco}
\end{figure}

\begin{figure}
    \centering
    \includegraphics[width=0.4\linewidth]{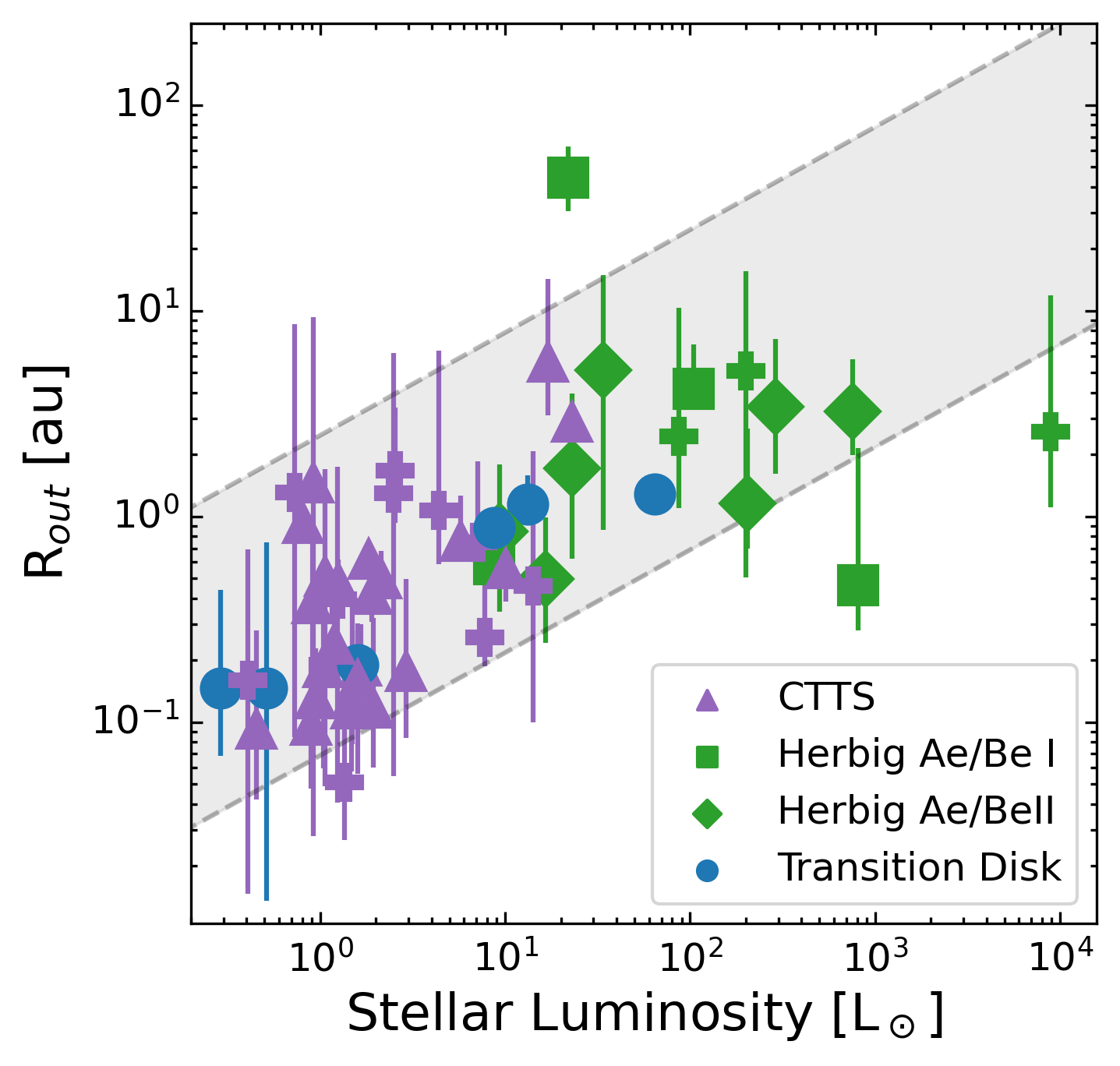}
    \includegraphics[width=0.4\linewidth]{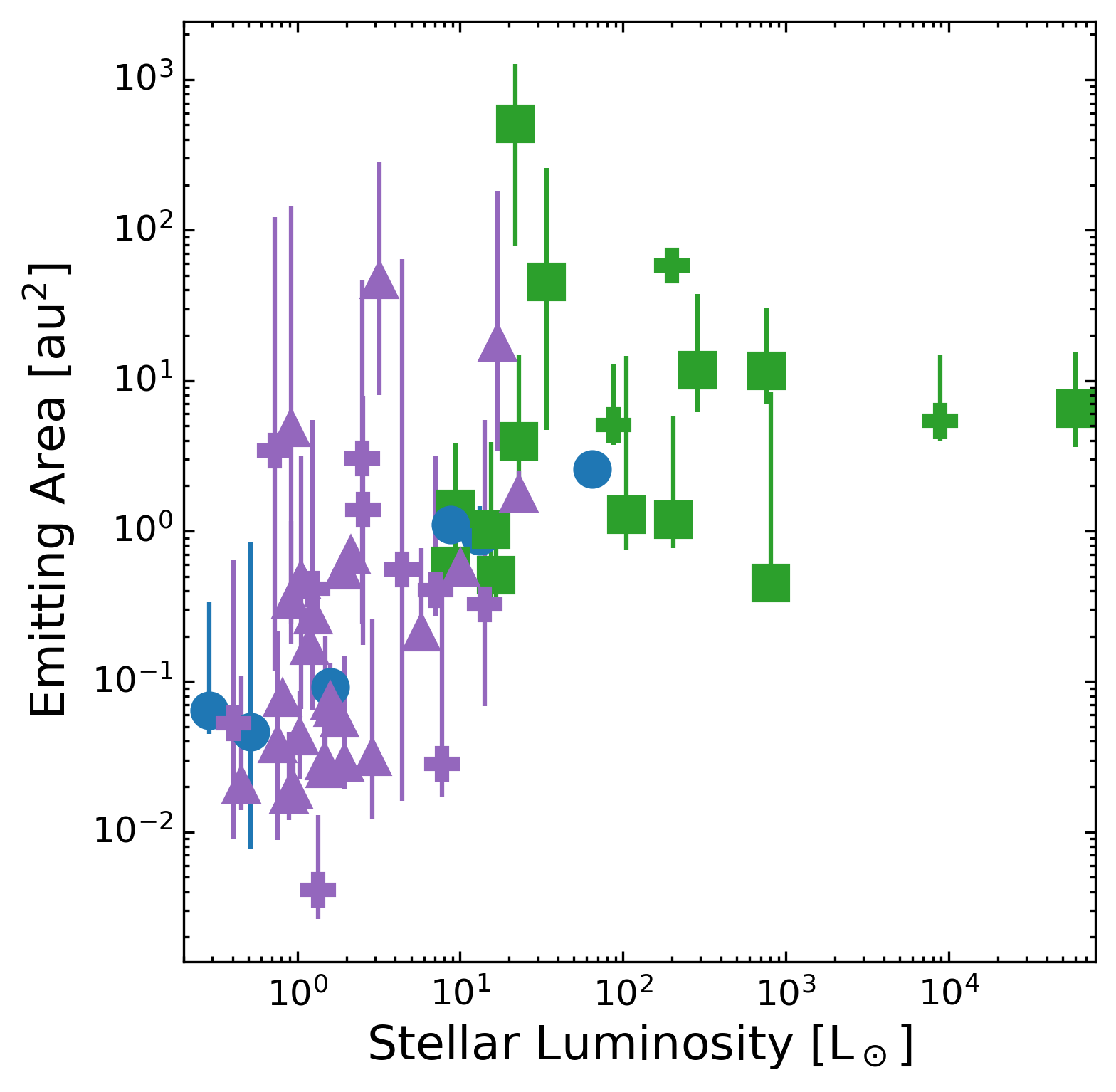}
    \caption{Left: Experimental radius versus stellar luminosity. The gray shaded region represents a range of excitation temperature radii cut offs, calculated from the stellar luminosity: the larger outer radii are for 250 K and the smaller outer radii for a temperature of 1500 K. Right: Retrieved Emitting Area versus stellar luminosity. Poor retrievals are marked with the plus sign, as defined in Section~\ref{subsec:SlabSpec}.}
    \label{fig:RoutLumi}
\end{figure}

\subsection{CO Emission Versus Accretion Tracers}\label{subsec:Herbigs}

Although the CO luminosity around both Herbigs and CTTS trends with accretion rate, accretion tracers and the presence of rovibrational CO are not as connected for Herbigs as they are for CTTS, as illustrated by Figure~\ref{fig:PBRates}. \citet{Ilee2018} suggests that the low CO overtone detection rate in Herbigs can be explained by a line-to-continuum contrast that makes observing CO only possible for stars with high accretion rates. Potentially, this also connects to the observability of CO fundamental emission. \citet{Banzatti2022} discusses that fundamental CO emission is a better tracer of molecular gas in these systems, since it can probe lower temperatures rather than the $>$1000 K excitation necessary for overtone emission. But, with systematically lower line-to-continuum ratios in Herbigs (Figure~\ref{fig:LTCfig}) despite their CO luminosities being similar to those of CTTS, rovibrational CO is still harder to detect in Herbigs than CTTS due to their stronger continuum emission. This is also consistent with many Herbig objects having only rovibrational CO detections in the lower energy levels, since the higher excitation levels have lower line fluxes and are buried under the bright continuum. 

There are a few possible explanations for why Herbigs seem to systematically lack rovibrational CO detections while still having accretion tracers: the gas in the inner disk is atomic due to photodissociation from the host star, the disk atmosphere of Herbigs is thinner and less settled \citep{Anotonellini2016}, or the gas from the disk is being funneled onto the star in small surface area streams despite a mostly cleared inner disk \citep{Casassus2013}. Although it remains ambiguous which is the prevailing cause for accretion tracers being present in disks without CO detections, further observations of photodissociation tracers and disk settling in Herbigs could help elucidate answers.

Also notable in Figure~\ref{fig:PBRates} is that the TTS show a trend opposite to the Herbigs: 18\% of the TTS sample has a CO emission detection and no Pf $\beta$, while only 6\% show the opposite. There are 7 objects that had CO emission detected but had poor wavelength coverage that excluded the Pfund $\beta$ region, but this does not account for the other 11 TTS sources. Ten of the twelve CTTS with no Pf $\beta$ signature are low accretors ($\log \dot{M}$ between -9 and -8 M$_\odot$ yr${-1}$). The remaining CTTS, V1331 Cyg, has a high accretion rate ($\log \dot{M} \sim$ -6.15) and very bright CO luminosities but there is no Pf $\beta$ present above 1$\sigma$. It is notably a pre-FU Ori system with complicated line veiling, which may factor into this non-detection \citep{Petrov2014}.

\begin{figure}
    \centering
    \includegraphics[width=.85\linewidth]{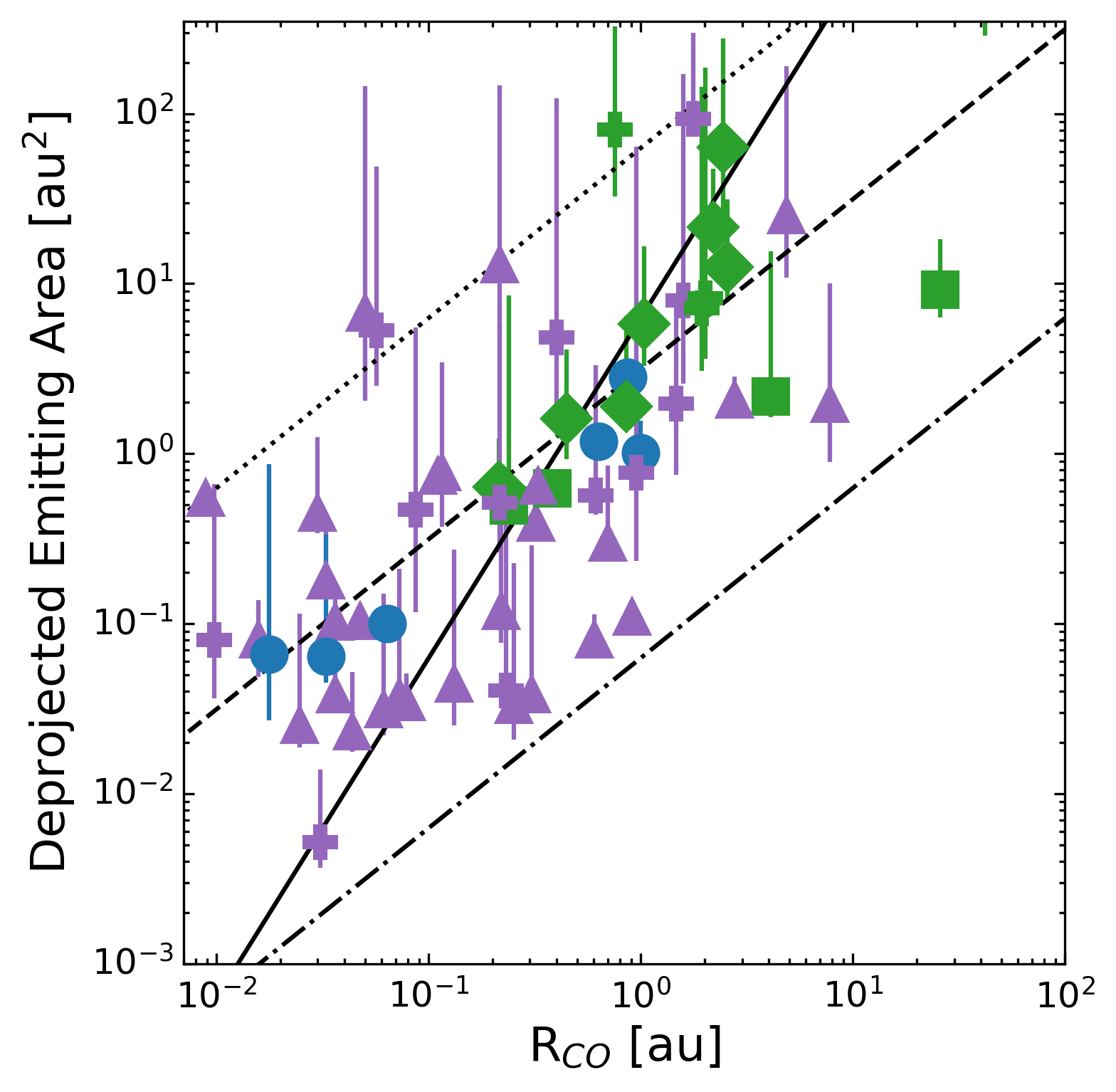}
    \vskip -0.1in
    \caption{The deprojected retrieved Emitting Area plotted against R$_{\rm CO}$, the Keplerian radius at HWHM. The purple triangles are CTTSs, the green squares are Herbigs, and the blue circles are transition disks, while objects with poor retrievals (as defined in Section~\ref{subsec:SlabSpec}) are marked with a plus symbol. The different dashed lines correspond to $A = 2\pi R\Delta R$: $\Delta R = 10$ is dotted, $\Delta R = 0.5$ is dashed, and $\Delta R = 0.01$ is dot-dashed. The solid black lines correspond to a circle, $A = 2\pi R^2$.}
    \label{fig:AreaRco}
\end{figure}

\begin{figure}
    \centering
    \includegraphics[width=0.31\linewidth]{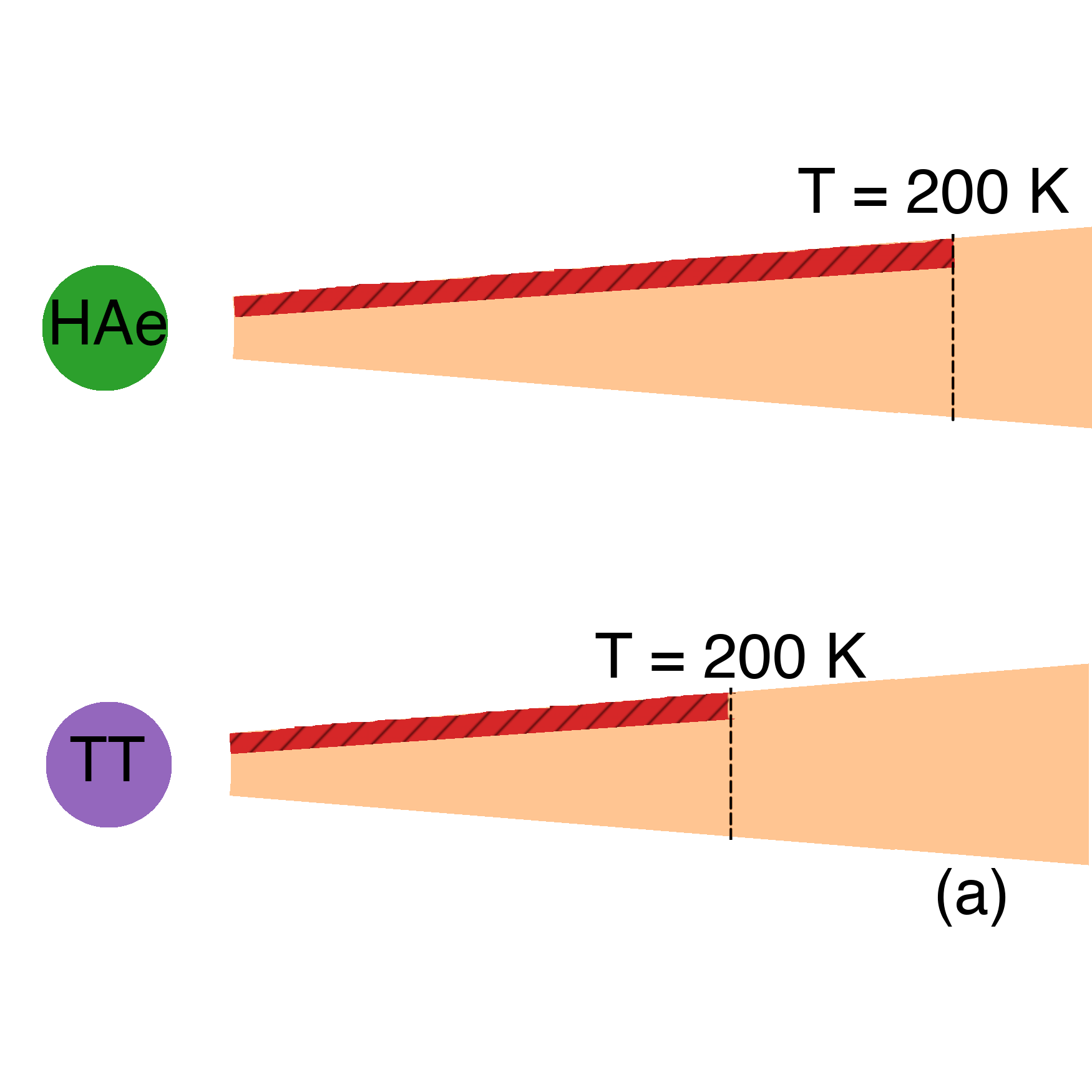}
    \includegraphics[width=0.31\linewidth]{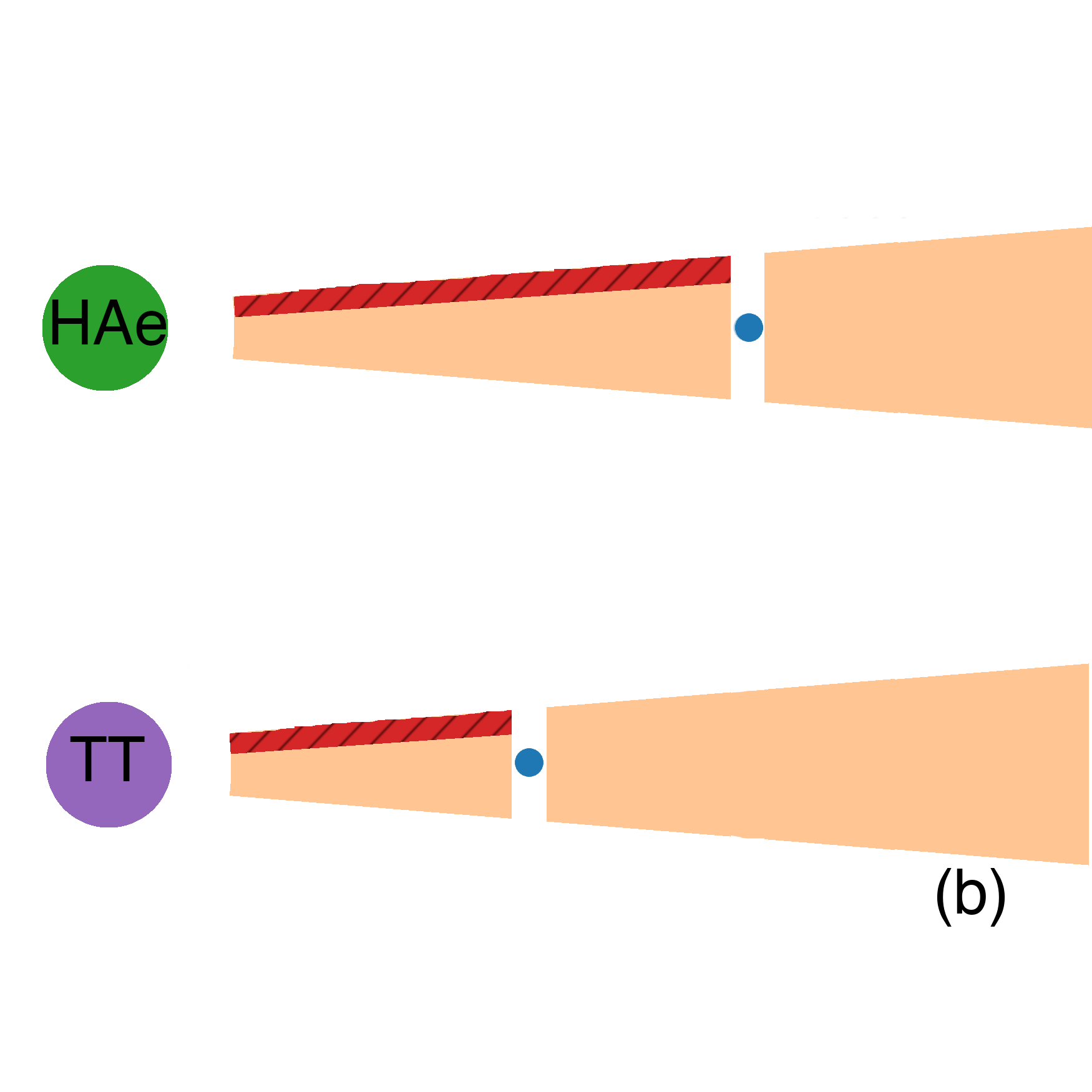}
    \includegraphics[width=0.31\linewidth]{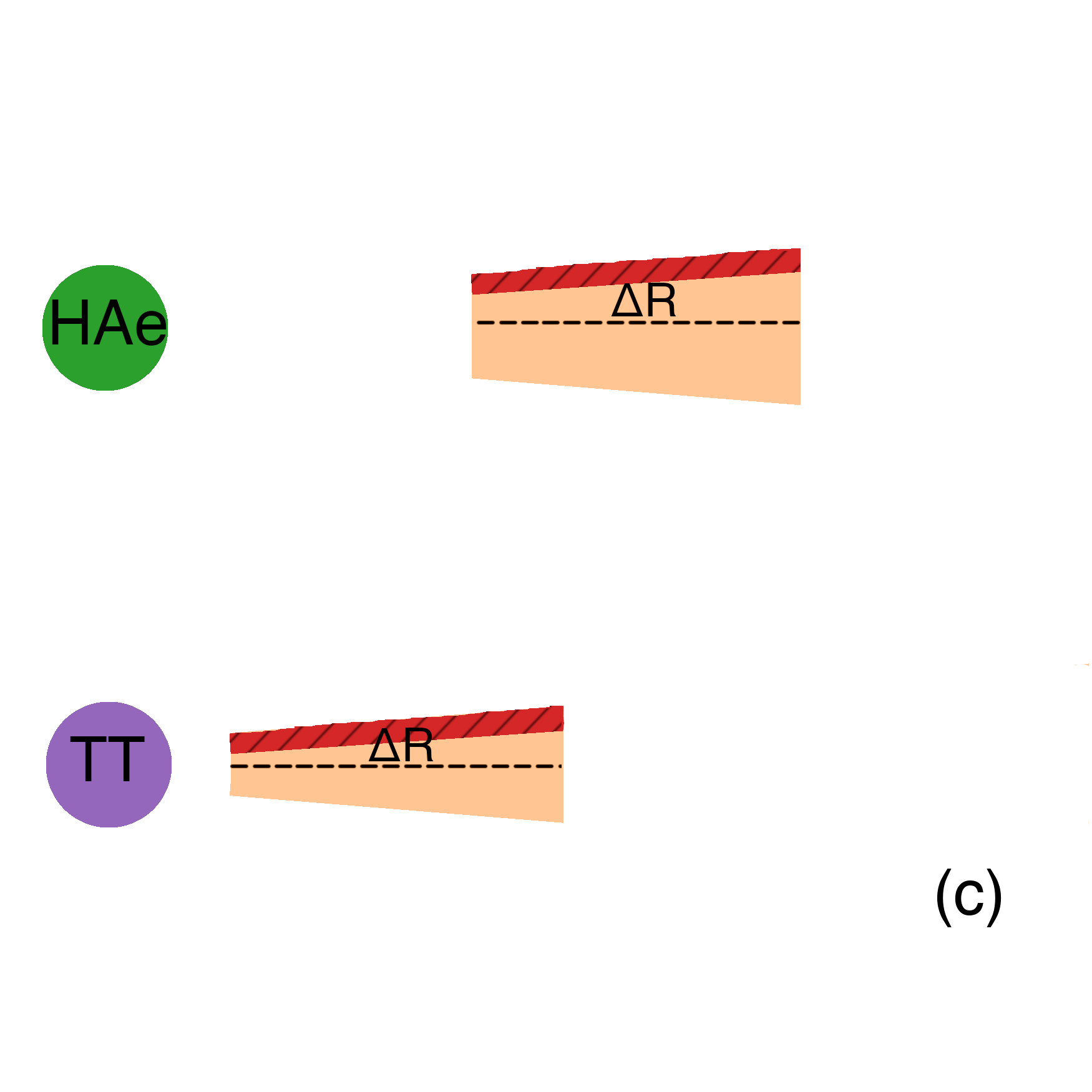}
    \vskip -0.25in
    \caption{The three different explanations for physical processes that could be causing the observed emitting area of the CO. In all three, the disk is represented in orange while the red, striped region represents the CO emitting area. \textit{(a)} The emitting area is determined by the temperature of the CO. \textit{(b)} The emitting area is determined by structure in the disk - in this cartoon by a planet-caused ring. \textit{(c)} The emitting area is always a thin ring of width $\Delta$R.}
    \label{fig:RadiiCartoon}
\end{figure}

\subsection{The {\rm CO} Emission Geometry}\label{subsec:EmitArea}

Our retrieved emitting areas and temperatures (Figure~\ref{fig:TempRco}) are consistent with a physical picture in which larger emitting radii arise from regions farther from the star \citep{Banzatti2022}. Double-peaked line shapes are associated with hotter stars (Herbigs) and interpreted as emission from inner disk rings, and are thought to come from a narrower emitting region, while triangular line shapes are associated with cooler stars (TTS) and are associated with disk conditions that are warmer out to much larger radii, perhaps due to lower physical densities or dust-to-gas ratios \citep{Banzatti2022}. Spectro-astrometric work has found that the triangular shaped emission could be sub-Keplerian and indicative of a wind instead of large Keplerian disk radii \citep{Pontoppidan2011}. 
We retrieve emitting areas for Herbigs that are larger overall than for TTSs (Fig~\ref{fig:RoutLumi}). This could be consistent with a narrow ring, as proposed in \citet{Banzatti2022}, if the Herbig objects consistently had larger inner radii. The TTSs in our sample have higher retrieved temperatures than the Herbigs, which is consistent with the more expansive emitting regions and warmer disk conditions predicted by the triangular line shapes associated with TTSs. Many of these TTS disks also have temperatures higher than the dust sublimation temperature, alluding to a significant portion of the rovibrational CO emitting from the inner dust-free region of the disk.

It does not appear that, overall, the CO emitting area can be estimated by a thin ring with constant $\Delta$R. Figure~\ref{fig:AreaRco} illustrates that while the emission shape is most likely that of a thin ring, rather than circular or nearly continuous, there is no single estimate of ring width that fits the observed positive trend between $R_{\rm CO}$ and emitting area. If we also consider the ratio of R$_{out}$ to R$_{CO}$, our sample shows a large range of values, from 1 to 50 implying a large range of ring thicknesses. While the inner radius might potentially trend with the retrieved temperatures of the gas (Figure~\ref{fig:TempRco}, left), the retrieved area {\em does} appear to negatively correlate with temperature (Figure~\ref{fig:TempRco}, middle). Similarly to the dust temperature relationship presented by \citet{Dullemond2001}, the temperature of the gas trends as (Radius)$^{-1/2}$. This implies that while both dust and gas temperature are directly linked to the stellar luminosity (Figure~\ref{fig:RoutLumi}), the processes caused by the stellar radiation affect the dust and gas differently, and therefore their temperatures are decoupled. 

In Figure~\ref{fig:RadiiCartoon}, we present three possible scenarios, including the thin ring, that could explain the retrieved emitting areas. The first scenario, $a$, imagines the CO emitting area as a function of temperature -- i.e., when the gas drops below a certain temperature then the gas would no longer be warm enough to emit. This means that the emitting area of the gas would trend positively with stellar luminosity as brighter stars can heat the disk to farther radii. Figure~\ref{fig:RoutLumi} does show that there is a difference between the outer radius and stellar luminosity trend in Herbigs and CTTS, but neither property trends with a sublimation radius linked to stellar radiation (the gray-shaded radius representing how radii would trend with stellar luminosity for a range of temperatures from 250 to 1500 K). CTTS's in particular seem to have a much broader range of outer radii than cutoffs based on stellar luminosity would predict. The Herbigs due tend to have larger outer radii than the CTTS, but there is still no trend with stellar luminosity. There is no evidence within our sample that the rovibrational CO emission emitting area is defined by a temperature cutoff.

The second scenario, $b$, proposes that instead the outer radius of the CO emitting area is determined by the structure of the disk, such as by a physical gap in the disk caused by the presence of a planet. If this were the case, we would expect a broad range of radii that would not be dependent on stellar luminosity, inner radius, or temperatures. While Figure~\ref{fig:AreaRco} does show that the inner radius positively correlates with the deprojected retrieved emitting area, there is too much dispersion within the trend to imply a single relation. As discussed in the previous paragraph, the outer radius is dependent on neither temperature nor stellar luminosity. The large dispersion in all of these correlations is suggestive of a scenario where the outer radii are determined by some other process.

Panel c of Figure~\ref{fig:RadiiCartoon} presents a final scenario, based on the results from \citet{Salyk2011b}, where the disk is always a thin ring of width $\Delta$R. But, as illustrated in Figure~\ref{fig:AreaRco}, the data are not in agreement with CO disk emission behaving consistently as a thin ring. There is some overall trend of increased emitting area with increased inner radius, but not a tight relation such as this scenario would predict.

Our results best agree with a scenario where the outer radii of the rovibrational CO emitting area is linked to the substructure of the inner disk. More sensitive measurements of disks where CO (2-1) and $^{13}$CO line fluxes can be used to better model the disks, as well as connections with other probes of disk structure (i.e. high resolution disk imaging), could further test the properties of the emitting regions' geometry.

\section{Lessons for JWST}\label{sec:JWST}

We present predictions for the observability of different objects from the line fluxes presented in our sample based on their stellar type (and accretion rate) in Table~\ref{table:LTC} for both the P(10) line (for JWST-NIRSPEC) and the P(30) line (for MIRI-MRS). The highest LTC ratios are seen in TTS, with a slight positive correlation with increased accretion rate, while Herbigs tend to have low LTC ratios independent of stellar luminosity or accretion. For Herbigs, Pf$\beta$ is more commonly observed than CO and therefore may be a more robust gas tracer, but CTTSs appear to be the better targets for studies of rovibrational CO and therefore may make more reliable targets for observing water and other molecular emission lines in the IR. 
Even with lower LTC ratios than CTTSs, Herbigs have brighter continuum emission which should allow for successful observations of CO with JWST-NIRSPEC (note the higher P(10) LTC ratio), but the higher excitation levels present in the JWST-MIRI wavelengths will be unlikely to be detected.
Of note, the 13 transition disks in our sample have a high P(10) LTC, while they have a lower corresponding P(30) LTC, falling into a similar pattern as the Herbigs.

\begin{figure}
    \centering
    \includegraphics[width=.9\linewidth]{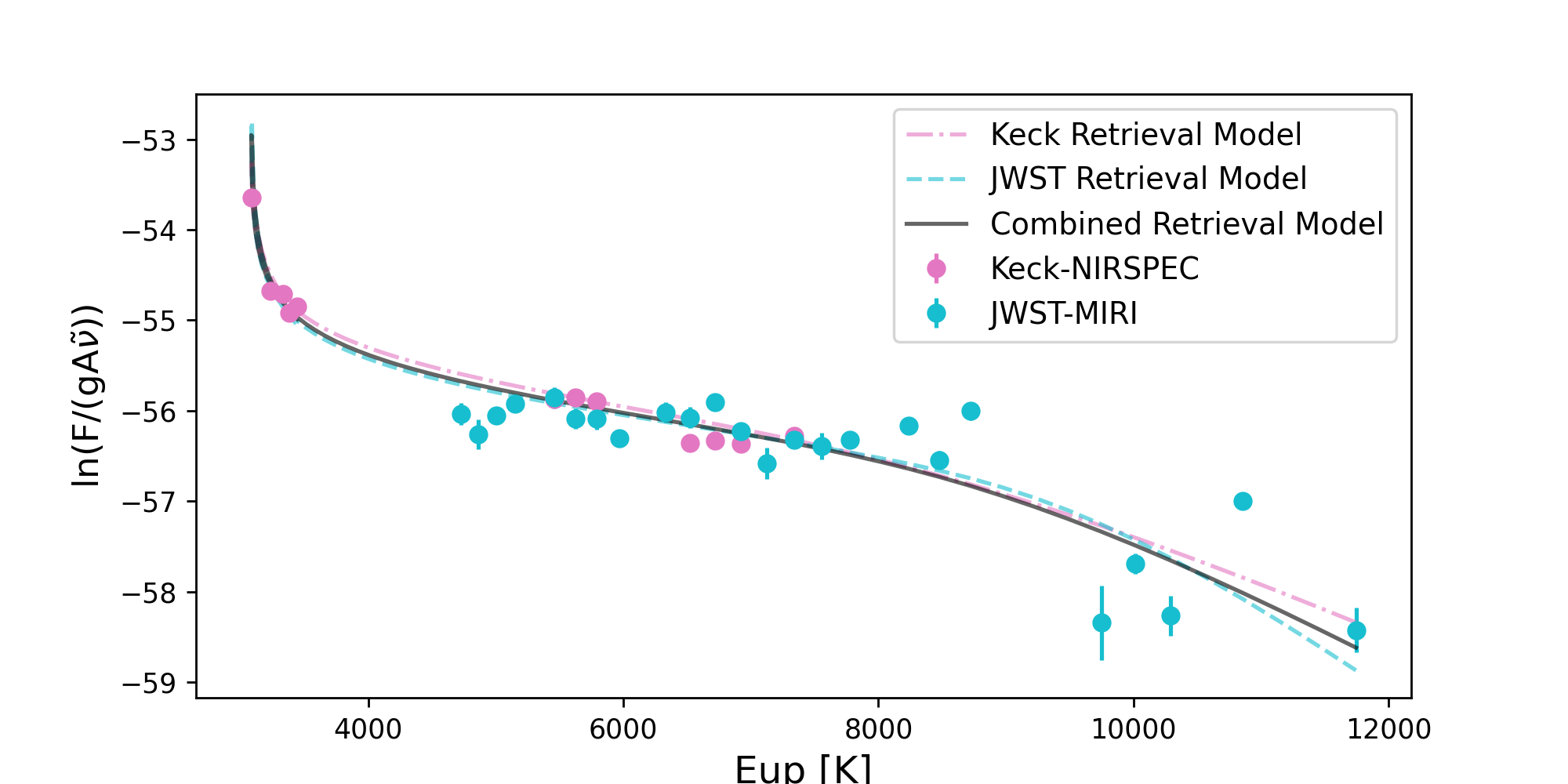}
    \caption{Rotation diagram for combined Keck-NIRPSEC and JWST-MIRI rovibrataional CO data for FZ Tau. Three slab models are shown on the graph based on three different data sets the retrieval was run on: just the Keck data (teal), just the JWST data (pink), and the combination of the two (black). }
    \label{fig:combined-data}
\end{figure}

While spectral observations with JWST have much higher sensitivity than previous ground-based capabilities, and excellent dynamic range driven by the stable instruments and lack of telluric features/variability, the spectral resolution is behind the capabilities of the best ground-based spectrographs (R$\sim$1000-3000 versus R$\sim$10000-100000). This means that information about the kinematics of the gas (as interpreted by the line shape) with JWST are limited; therefore the detailed line width and line shape analysis is not able to be performed on these spectra \citep[i.e.,][]{Banzatti2022}. 

With the addition of high-resolution ground-based NIR spectra, \citet{Temmink2024a} was able to greatly expand their analysis of their MIRI-MRS spectrum of DR Tau. They found that the excitation temperature derived from just the MIRI data was significantly higher than with the combined data. With the better spectrally-resolved ground-based spectra they were able to properly characterize complex line profiles and then apply the information gained from the line shape to the lower resolution MIRI data.

Even without the higher spectral resolutions needed to analyze the line shape properties, combining ground-based data with a MIRI spectrum expands the excitation level coverage for more informed modeling and, therefore better, parameters for the inner disk CO reservoir. The part of the P-series covered by MIRI MRS is at high upper energy levels, which is biased towards higher excitation temperatures. The earlier part of the P-series with a broader range of upper energy levels, covered by ground-based instruments and JWST-NIRSpec, is more sensitive to optical depth effects and lower excitation temperatures. 

We combined a spectrum from MIRI-MRS with ground-based spectra to create a ``full'' coverage P-series rotation diagram for FZ Tau with data from this study and \citet{Pontoppidan2023} (Figure~\ref{fig:combined-data}). While the ground-based Keck-NIRSPEC data extend from P(2) -- P(40), with many gaps in that coverage, the MIRI spectrum extends to higher energy transitions: P(25) -- P(57). We did three sets of retrieval slab modelling on FZ Tau: just the Keck-NIRSPEC emission lines, just the JWST-MIRI emission lines, and the two sets of emission lines combined. While the retrieved slab models do not have column densities that vary significantly ($\sim$10$^{18}$ cm$^{-2}$ for all solutions), the temperature retrieved varies greatly: 1600 K (just Keck), 1000 K (just MIRI), to 1300 K (combined Keck and MIRI). The higher energy transitions that are visible in the MIRI wavelength coverage are more sensitive to the temperature of these objects; combined slab modeling on these objects will better probe the temperatures of these objects. 

For TTSs, JWST-MIRI data can be easily utilized to better constrain temperatures, as illustrated in Figure~\ref{fig:combined-data} for FZ Tau.  The combined rotation diagram probes farther energy levels which allow for better retrieval fits. The JWST-MIRI data also expands the middle energy levels due to the lack of atmospheric bands requiring specific wavelength ranges to be ``dropped'' from ground-based observing. 

\begin{figure}
    \centering
    \includegraphics[width=0.9\linewidth]{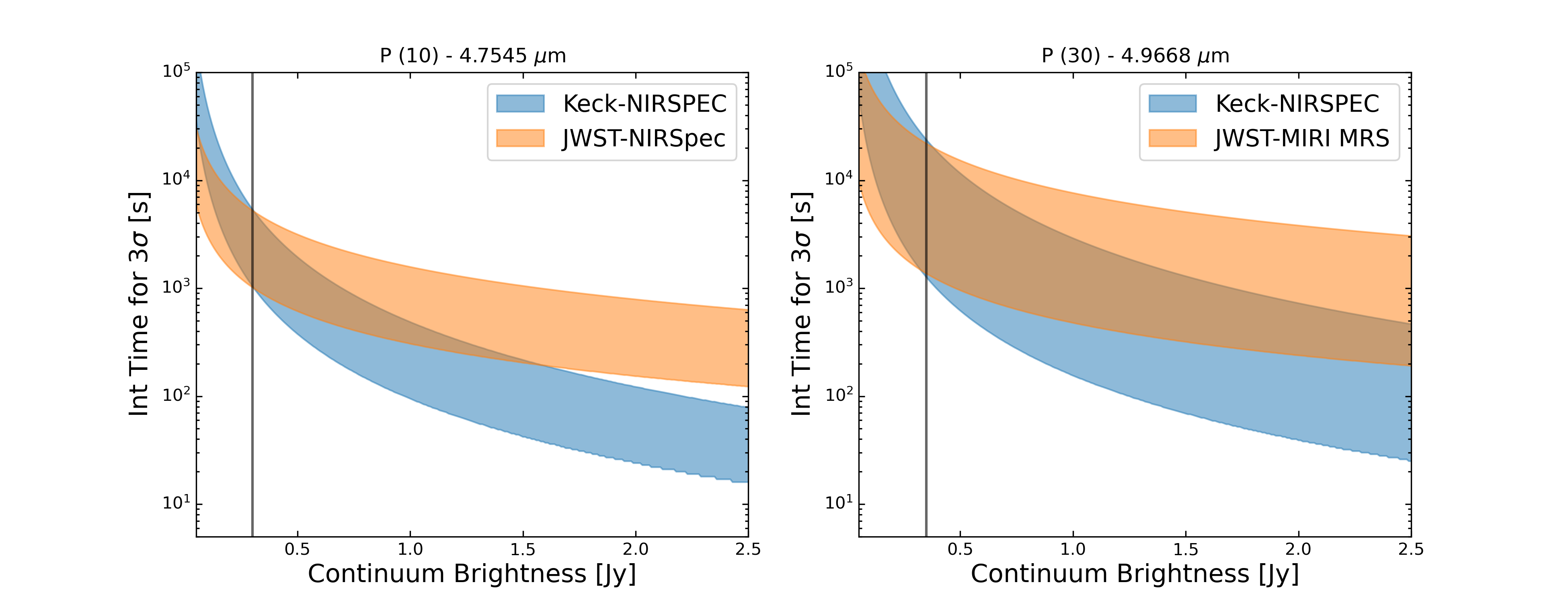}
    \caption{Left: Integration time to reach a 3$\sigma$ detection of the P(10) CO rovibrational emission line versus the continuum flux of the object at 4.75 $\mu$m. The shaded region correspond to a range of LTC ratios (0.22 to 0.50 for Keck, and 0.11 to 0.25 for JWST), representative of average values from our sample. The intersection of Keck and JWST is represented by a gray line at 0.3 Jy. Right: Same as the left, but for the P(30) CO rovibrational emission line and JWST-MIRI MRS and the continuum flux at 4.95 $\mu$m. The range of LTC ratios represented for the P(30) line are 0.09 to 0.39 for Keck, and 0.05 to 0.20 for JWST. The intersection line is at 1.03 Jy.}
    \label{fig:IntTimes}
\end{figure}

Figure~\ref{fig:IntTimes} shows the predicted integration times needed for a 3$\sigma$ detection of rovibrational CO at 4.5 $\mu$m with Keck-NIRSPEC versus JWST-NIRSpec. These integration times were calculated to achieve the correct S/N ratio (ranging from 7 to 66) for the appropriate LTC ratios (ranging from 0.05 to 0.5). We assumed that the JWST noise was photon-limited while the Keck noise was background-limited and adjusted the integration times by anchoring the numbers to real data. We found that the integration time as a function of continuum brightness followed 
\begin{equation}
    t = A L^{b}
\end{equation}
where t is the integration time in seconds and L is the source brightness in Jy. For JWST, b was equal to -1 and for Keck-NIRSPEC, b was equal to -2. The value used for A was dependent on the desired S/N ratio squared. 

We predict that JWST-NIRSpec will be able to detect rovibrational CO with reasonable integration times (under an hour) in faint objects ($<$0.3 Jy) that are too intensive to observe from the ground, even with the decreased LTC ratios due to lower spectral resolution. The low excitation level CO lines present in JWST-NIRSPEC have higher LTC ratios overall and make even faint objects good targets for possible CO detections. For the longer wavelengths, JWST-MIRI is almost always better at detecting rovibrational CO due to the much higher S/N capabilities and the lower LTC ratios. For most objects with a lack of, or marginal, CO detection in a JWST-MIRI spectrum, follow-up ground-based high resolution spectra will be able to detect CO easily. 

Especially in the case of Herbig disks (which are on average cooler disks), non-detection of CO does not mean that there is no CO present in the inner disk. As discussed in Section~\ref{subsec:LF}, Herbigs have lower LTC ratios and lower CO temperatures which makes rovibrational CO harder to detect in these systems. Only the lowest energy levels of the rovibrational emission will be observable even in JWST-NIRSpec wavelengths. The non-detections can be used in combination with NIR low P observations of CO (either via ground-based spectra or JWST-NIRPSEC) to constrain temperatures of these disks. Although JWST is more sensitive to these objects, the emission lines will not be resolved due to the lower resolution of its spectrometers and will have lower line-to-continuum ratios (see Table~\ref{table:LTC}) that may still get buried in the noise. 

For JWST data that has limited to no CO rovibrational line information, i.e. data from just JWST-MIRI, the retrieved gas properties presented in Table~\ref{table:RetrievedProps} can be used in conjunction with other molecular line analysis to constrain molecular ratios of molecules observed, such as H$_2$O. For Herbigs especially, it is less likely there will be sufficient CO rovibrational lines when observing only the high upper energy levels observable with JWST-MIRI.

\section{Summary and Conclusions}\label{sec:Summ}

The goal of this work has been to provide a comprehensive reference of two decades of spectra taken by Keck-NIRSPEC. We have analyzed each spectrum for the presence of rovibrational CO emission lines and Pf$\beta$ and followed up any detections with slab modeling. The most notable conclusions of this work are:

\begin{enumerate}
    \item We analyze 183 Keck-NIRSPEC M-band spectra of young stars with disks. These spectra represent $\sim$60\% of known disks that fall within Keck-NIRSPEC's observing limitations. The reduced spectra are available to the public via the {\tt SpExoDisks} Database. We detect rovibrational CO emission lines in 96 of these objects.
    \item In both WTTS and Class III objects, we detect no rovibrational CO emission and CO absorption in only one of the WTTS. These objects have neither accretion nor molecular gas detections, proving that their inner disks are most likely completely cleared of material in the gas-phase. We find that CO is a reliable tracer of gas in the inner disks of CTTS, while less so for Herbigs due to high accretion rates potentially dissociating the CO.
    \item We modeled via slab modeling the gas for 67 objects with sufficient CO detections and present those in Table~\ref{table:RetrievedProps} and Figure~\ref{fig:RetResults}. We provide guidance on how to use these slab model parameters for better analysis of future JWST data. 
    We find that TTS disks have rotation diagrams consistent with higher temperatures and smaller emitting areas as compared to Herbigs. This is consistent with a picture in which the CO in T Tauri disks arises from smaller disk radii. Column densities are similar for the two types of disks.
    \item The CO luminosity is positively correlated with the accretion rate for the full sample (Figure~\ref{fig:lumaccretion}). This implies that the temperatures of the molecular CO inner disk reservoirs are connected with the ongoing accretion in these systems. This correlation is again seen in the retrieved temperatures, but only in CTTS. Herbigs have CO luminosities that correlate with accretion rate but not temperature, implying that either at higher mass accretion rates these two processes become untethered or at greater stellar luminosities accretion has less effect on line flux. 
    \item The retrieved areas for the disks do trend positively with R$_{\rm CO}$,  but not as a thin ring with standard $\Delta$R. We present two additional alternative models for the emitting area of CO (Figure~\ref{fig:RadiiCartoon}). We find no dependence of the outer radii on stellar properties, which implies that the substructure of the inner disk, i.e. a gap in the disk, is the cause for dispersion seen in the outer radii derived from the CO emitting area.
    \item  We present guidelines on how to use ground-based CO emission lines in combination with JWST data in Sec.~\ref{sec:JWST}. We suggest that CTTS are better targets for probing rovibrational CO lines using JWST due to brighter line flux-to-continuum ratios, even with the increased sensitivity of JWST. We predict LTC ratios for differing mass and accretion properties and present them in Table~\ref{table:LTC}. We estimate that JWST will be more suited for observing CO in fainter disks than from ground-based telescopes, especially low-mass stars with high accretion rates (which have higher predicted LTC ratios), but worse suited for Herbigs (which have lower predicted LTC ratios).
\end{enumerate}

\begin{acknowledgements}
We thank Andrea Banzatii, Joan Najita, and Sebastiaan Krijt for helpful discussion that aided in the science of this paper. Additionally, we thank the anonymous referee for insightful comments that improved this manuscript. This work was supported by grant \# 2107445 from the National Science Foundation, entitled "RUI: Characterizing Inner Disks in the era of JWST and ALMA". A portion of this research was carried out at the Jet Propulsion Laboratory, California Institute of Technology, under a contract with the National Aeronautics and Space Administration (80NM0018D0004).
\end{acknowledgements}

\bibliography{main}
\bibliographystyle{aasjournal}

\begin{center}
    \small
\begin{longrotatetable}
\begin{deluxetable*}{lccccccccccr}
\tablecaption{The Sample: Literature Values}
\label{table:Sample}
\tablehead{\colhead{Name} & \colhead{Detections$^a$} & \colhead{Distance$^b$} & \colhead{SpT} & \colhead{M$_*$$^c$} & \colhead{$\log{L_*}$} & \colhead{Class$^d$} & \colhead{Type} & \colhead{Subtype$^e$} & \colhead{$\log$M$_{acc}$} & \colhead{Incl}& \colhead{REFS}\\
\colhead{} & \colhead{} & \colhead{[pc]} & \colhead{} & \colhead{[M$_\odot$]} & \colhead{[$\log{L_\odot}$]} & \colhead{} & \colhead{} & \colhead{} & \colhead{M$_\odot$yr$^{-1}$} & \colhead{$^{\circ}$} & \colhead{}}
\startdata
49 Cet & \nodata & 57.11 & A1 & \nodata & \nodata & d & Debris & \nodata & \nodata & 79.3$^{+0.4}_{-0.4}$ & \ref{BJ21},\ref{H88},\ref{Z12},\ref{Hu17} \\
51 Oph & \nodata & 125.26 & A0 & \nodata & \nodata & d & Debris & \nodata & \nodata & 63$^{+1}_{-1}$& \ref{BJ21},\ref{D97},\ref{G21} \\
AA Tau & CO & 133.69 & M0.6 & 0.486 & -0.12 & II & T Tauri & C & -8.37 & 30$^{+8}_{-7}$ & \ref{BJ21},\ref{HH14},\ref{PP7},\ref{F06},\ref{Betti23},\ref{BE19} \\
AB Aur & CO,13,21,PB & 155.94 & A1 & 2.56 & 1.81 & T & Herbig Ae/Be & I & -6.8 & 23 & \ref{BJ21},\ref{HH14},\ref{M13},\ref{FM20},\ref{GD21},\ref{vM16},\ref{GL06},\ref{T12} \\
AS 205 N & CO,13,21,PB & 131.72 & K5 & 0.87 & 0.33 & II & T Tauri & C & -7.63 & 20.1$^{+3.3}_{-3.3}$ & \ref{BJ21},\ref{C79},\ref{GD21},\ref{L96},\ref{Betti23},\ref{K18} \\
AS 205 S & \nodata & 140.48 & M3 & 0.3 & \nodata & II & T Tauri & C & -7.6 & 66.3$^{+1.7}_{-1.7}$ &\ref{BJ21},\ref{C79},\ref{L96},\ref{Betti23},\ref{K18} \\
AS 209 & CO,PB & 120.74 & K4 & 0.6$^\dagger$  & \nodata & \nodata & T Tauri & \nodata & \nodata & \nodata & \ref{BJ21},\ref{T06} \\
AS 353 A & \nodata & 377.67 & K1 & 1.6 & \nodata & II & T Tauri & C & -7.1 & 41$^{+2}_{-2}$ &\ref{BJ21},\ref{Rei18},\ref{G21} \\
\enddata
\end{deluxetable*}
\end{longrotatetable}
\end{center}

\begin{center}
    \begin{deluxetable*}{lcccccccc}
\tablecaption{Line Fluxes}
\label{table:Linefluxes}
\tablehead{\colhead{Name} & \colhead{R(2)} & \colhead{R(1)} & \colhead{R(0)} & \colhead{P(1)} & \colhead{P(2)}  & \colhead{--}& \colhead{P(39)} & \colhead{P(40)}} 
\startdata
AA Tau & \nodata & \nodata & 1.909e-17 & 9.638e-18 & 1.411e-17  & -- &\nodata & 1.086e-17 \\
AB Aur & 1.028e-16 & 9.227e-17 & \nodata & \nodata & \nodata & -- & 1.998e-17 & 1.920e-17 \\
AS 205 N & \nodata & \nodata & \nodata & \nodata & 1.696e-16 & -- & \nodata & 1.175e-16 \\
BP Tau & \nodata & \nodata & \nodata & \nodata & \nodata & -- & \nodata & \nodata \\
CI Tau & \nodata & \nodata & \nodata & 1.033e-16 & 6.637e-17 & -- & \nodata & 8.449e-17 \\
\enddata
\tablecomments{Line Fluxes are in W m$^{-2}$. The full version of this table, including information on all transitions, is available in machine readable format.}
\end{deluxetable*}
\end{center}

\begin{center}
    \startlongtable
\begin{deluxetable*}{lcccccc}
\tablecaption{Retrieved and Derived Properties of the Sample}
\label{table:RetrievedProps}
\tablehead{\colhead{Name} & \colhead{Log Column Density} & \colhead{Temperature} & \colhead{Emitting Area} & \colhead{R$_{ret}$} & \colhead{R$_{out}$} & \colhead{Flags} \\\colhead{} & \colhead{cm$^2$} & \colhead{K} & \colhead{au$^2$} & \colhead{au} & \colhead{au} & \colhead{} }
\startdata
AA Tau & 19.04$^{+0.67}_{-0.75}$ & 1289$^{+843}_{-508}$ & 0.04$^{+0.18}_{-0.03}$ & 0.12$^{+0.26}_{-0.10}$ & \nodata & T \\
AB Aur & 17.44$^{0.54}_{-0.06}$ & 874$^{+26}_{-39}$ & 2.59$^{+0.32}_{-0.2}$ & 0.95$^{+0.33}_{-0.26}$ & 1.28$^{+0.33}_{-0.26}$ & \nodata \\
AS 205 N & 17.86$^{+0.08}_{-0.06}$ & 1284$^{+63}_{-70}$ & 0.70$^{+0.10}_{-0.07}$ & 0.49$^{+0.18}_{-0.16}$ & 0.50$^{+0.18}_{-0.16}$ & \nodata \\
AS 209 & 19.68$^{0+.24}_{-0.57}$ & 1940$^{+756}_{-875}$ & 0.02$^{+0.19}_{-0.01}$ & 0.10$^{+0.30}_{-0.08}$ & 0.27$^{+0.30}_{-0.08}$ & T \\
CI Tau & 18.70$^{+0.18}_{-0.2}$ & 2704$^{+210}_{-343}$ & 0.07$^{+0.03}_{-0.01}$ & 0.18$^{+0.11}_{-0.07}$ & 0.19$^{+0.11}_{-0.07}$ & \nodata \\
CW Tau & 16.71$^{+1.6}_{-1.2}$ & 2589$^{+263}_{-1233}$ & 0.48$^{+2.7}_{-0.41}$ & 0.50$^{+1.2}_{-0.46}$ & 0.51$^{+1.2}_{-0.46}$ & T \\
DF Tau & 19.22$^{+0.46}_{-0.42}$ & 1397$^{+207}_{-455}$ & 0.35$^{+0.81}_{-0.11}$ & 0.38$^{+0.57}_{-0.21}$ & 0.38$^{+0.57}_{-0.21}$ & \nodata \\
DG Tau & 15.85$^{+0.80}_{-0.80}$ & 1128$^{+186}_{-107}$ & 46.5$^{+240}_{-39}$ & 4.75$^{+11.0}_{-4.3}$ & \nodata & \nodata \\
\enddata

\tablecomments{F indicates the objects have retrieval results that were indicative of degenerative column density and emitting area values. T indicates that the objects had poorly constrained temperatures.}
\end{deluxetable*}
\end{center}

\begin{center}
    \startlongtable
\begin{deluxetable*}{lcccc}
\tablecaption{Line Profile Parameters}
\label{table:LineProfs}
\tablehead{\colhead{Name$^a$} & \colhead{Lines in Composite} & \colhead{Average Linewidth} & \colhead{Stacked FWHM} & \colhead{R$_{CO}$$^b$} \\ \colhead{}  & \colhead{}  & \colhead{km s$^{-1}$} & \colhead{km s$^{-1}$} & \colhead{au} }
\startdata
AA Tau$^*$  & P [30, 12, 11, 8, 7, 6, 5, 1] & 186$\pm$99 & \nodata & \nodata \\ 
AB Aur & P [38, 37, 36, 31, 30, 28, 26, 14, 12, 11, 8, 7, 6, 5] & 25$\pm$1 & 23 & 0.87 \\ 
AS 205 N & P [47, 45, 43, 37, 36, 31, 30, 14, 12, 11, 8, 6, 5] & 36$\pm$2 & 33 & 0.11 \\ 
BP Tau & P [37, 36, 12, 11, 8, 6] & 100$\pm$9 & 97 & 0.02 \\ 
CI Tau & P [30, 12, 11, 8, 7, 6, 5] & 121$\pm$7 & 99 & 0.05 \\ 
CW Tau$^*$ & P [36, 30, 12, 11, 8, 6, 5] & 115$\pm$93 & 86 & 0.12 \\ 
CY Tau & P [36, 30, 12, 8, 7] & 74$\pm$16 & 64 & 0.09 \\ 
DE Tau & P [37, 36, 12, 11, 8, 7] & 81$\pm$17 & 65 & 0.03 \\
\enddata
\tablecomments{a) $^*$ after a name indicates an object with a double-peaked profile. \\ b) $^\dagger$ after a R$_{CO}$ value indicates that no literature inclination was found for this object and therefore an inclination of 45$^\circ$ was used.}
\end{deluxetable*}
\end{center}

\begin{center}
    \begin{deluxetable*}{lccccc}
    \tablecaption{Mean LTC Values at Keck and JWST Resolutions}
    \label{table:LTC}
    \hspace{2em}
    \tablehead{\colhead{Type} & \colhead{log M$_{acc}$} & \colhead{P(10)} & \colhead{} & \colhead{P(30)} & \colhead{} \\
    \colhead{} & \colhead{[M$_\odot$ yr$^{-1}$]} & \colhead{Keck} & \colhead{JWST} & \colhead{Keck} & \colhead{JWST}}
    \startdata
    CTTS & [-9.0, -8.2] & 0.391$\pm$0.16 & 0.158$\pm$0.08 & 0.255$\pm$0.07 & 0.148$\pm$0.03\\
    CTTS & [-8.2, -7.4] & 0.499$\pm$0.23 & 0.250$\pm$0.12 & 0.392$\pm$0.23 & 0.230$\pm$0.12\\
    CTTS & [-7.4, -6.6] & 0.601$\pm$0.48 & 0.306$\pm$0.19 & 0.46$\pm$0.34 & 0.276$\pm$0.15\\
    Herbigs & all & 0.219$\pm$0.10 & 0.140$\pm$0.08 & 0.094$\pm$0.05 & 0.054$\pm$0.03\\
    TDs & all & 0.573$\pm$0.46 & 0.573$\pm$0.44 & 0.262$\pm$0.16 & 0.152$\pm$0.13
    \enddata
    \end{deluxetable*}
\end{center}

\end{document}